\documentclass[12pt]{article}
\usepackage[]{graphicx}
\usepackage{makeidx}
\usepackage{amsmath}
\usepackage{amssymb}
\usepackage{wasysym}
\usepackage{xcolor}
\usepackage{hyperref}
\usepackage{marvosym}
\usepackage[cm]{fullpage}
\hypersetup{citebordercolor=0 1 0, linkbordercolor=1 1 1,}

\newcommand{\be}{\begin{eqnarray}}
\newcommand{\ee}{\end{eqnarray}}
\newcommand{\rar}{\rightarrow}

\begin{document}

\begin{titlepage}
\title{Relic gravitational waves from light primordial black holes}
\author{Alexander D. Dolgov$^{a,b,c}$, Damian Ejlli$^{a,b}$}
\maketitle
\begin{center}
$^{a}$Universit\`a degli Studi di Ferrara, I-44100 Ferrara, Italy \\
$^{b}$INFN, Sezione di Ferrara, I-44100 Ferrara, Italy \\
$^{c}$ITEP, 113259 Moscow, Russia 
\end{center}
\thispagestyle{empty}
\date{}

\begin{abstract}

The energy density of relic gravitational waves (GWs) emitted by primordial black holes (PBHs) is 
calculated. We estimate the intensity of GWs produced at quantum and classical scattering of
PBHs, the classical graviton emission from the PBH binaries in the early Universe, and the graviton 
emission due to PBH evaporation. If nonrelativistic PBHs dominated the cosmological energy 
density prior to their evaporation, the probability of formation of dense clusters of PBHs and their 
binaries in such clusters would be significant and the energy density of the generated gravitational 
waves in the present day universe could exceed that produced by other known mechanisms. The 
intensity of these gravitational waves would be maximal in the GHz frequency band of the spectrum 
or higher and makes their observation very difficult by present detectors but also gives a rather good possibility to investigate it by present and future high frequency gravitational waves electromagnetic detectors. However, the low 
frequency part of the spectrum in the range $f\sim 0.1-10$ Hz may be detectable by the planned 
space interferometers DECIGO/BBO. 

For sufficiently long duration of the PBH matter dominated stage the cosmological energy fraction of 
GWs from inflation would be noticeably diluted.

\end{abstract}

\vspace{5cm}
\Email{Alexander D. Dolgov:  \href{mailto:dolgov@fe.infn.it}{\nolinkurl{dolgov@fe.infn.it}} }\\

\Email{Damian Ejlli: \href{mailto:ejlli@fe.infn.it}{\nolinkurl{ejlli@fe.infn.it}}}

\end{titlepage}

\section{Introduction}
\pagenumbering{arabic}

Since the prediction of gravitational waves (GW) by Albert Einstein in 1916~\cite{Einstein:1916}
on the basis of general relativity, they have been 
an object of intensive studies. Gravitational
waves are thought to be fluctuations in the curvature of space-time,
which propagate as  waves, traveling outward from the source. 
Although gravitational radiation has not yet been directly detected, it has been indirectly shown to exist
because it increases the pulsar orbital frequency \cite{Hulse:1974eb} 
in good agreement with theoretical predictions.

Roughly speaking there are two groups of
possible sources of gravitational radiation which may be registered by
gravitational wave detectors either on the Earth or by space missions. The first group includes
energetic phenomena in the contemporary universe, such as emission of GWs by black hole or
compact star binaries, supernova explosions, and possibly some other catastrophic phenomena.
The second group contains gravitational radiation coming from the early Universe, which
creates today an isotropic background usually with rather low frequency.
Such gravitational radiation could be produced at inflation, phase transitions in the primeval plasma, 
by the decay or interaction of topological defects, e.g. cosmic
strings, {\it etc}. 

The graviton (gravitational wave) production in the Friedmann-Robertson-Walker metric was first considered
by Grishchuk~\cite{Grishchuk:1974ny}, who
noticed that the graviton wave equation is not conformal invariant and thus such quanta can
be produced by conformal flat external gravitational field. Generation of gravitational waves 
at the De Sitter (inflationary)
stage was studied by Starobinsky~\cite{Starobinsky:1979ty} (see also ref.~\cite{Rubakov:1982df}). 
The stochastic homogeneous background of the low frequency gravitational waves is 
now one of the very important predictions of inflationary cosmology, which may present a final proof of
inflation. 

In this work we discuss one more source of gravitational wave (GW) radiation in the early
Universe, namely, the interaction between primordial black holes (PBH). We consider 
relatively light PBH, such that they evaporated before the big bang
nucleosynthesis (BBN) and so 
they are not constrained by the light element abundances. 
Cosmological scenario with early formed and evaporated primordial
black holes producing gravitons was considered
in ref.~\cite{Dolgov:2000ht}. Here we will remain in essentially the same frameworks and
study in addition the GW emission in different processes with PBH.

According to ref.~\cite{Hawking:1976de, DonPage:1976} the life-time of evaporating black hole
with initial mass $M$ is equal to: 
\be 
\tau_{BH} = \frac{10240\,\pi}{N_{eff}}\, \frac{M^3}{m_{Pl}^4}\,,
\label{tau-BH}
\ee
where the Planck mass is $m_{Pl}=2.176\cdot 10^{-5}$ g and
$N_{eff}$ is the number of particle species with masses
smaller than the black hole temperature: 
\be
T_{BH}=\frac{m_{Pl}^2}{8\pi M}\, .
\label{T-BH}
\ee
To avoid a conflict with BBN the black holes should had been evaporated 
before cosmological time $t\approx 10^{-2}$ s~\cite{Carr:2009jm} and thus their mass would be bounded from above by
\be
M < 1.75\cdot 10^8 \left(\frac{N_{eff}}{100}\right)^{1/3}\,\,{\rm g}.
\label{M-upper-bound}
\ee
The temperature of such PBHs should be higher than $3\cdot 10^4 $ GeV and
correspondingly $N_{eff} \geq 10^2$.
On the other hand, as is discussed in what follows, 
the PBH mass is bounded from below e.g. by equation (\ref{M-large-Delta}). 
This is the mass range of PBHs considered in this work. Such PBH are not
constrained by any astronomical data, which are applicable to heavier
ones~\cite{Carr:2009jm}--\cite{Josan:2009qn}.

Primordial black holes should interact in the early Universe creating gravitational radiation. 
Below we estimate the  efficiency of GW emission in 
several processes with PBH. In sec.~\ref{s-BH-prod}  some
mechanisms of PBH production and PBH evolution in the early
Universe are briefly described. We stress, in particular, a very important role played by
the clumping of PBH due to gravitational instability at the  
matter dominated stage. 
In section \ref{s-onset} we consider the initial 
interaction between the PBHs when they started to "feel" each other and accelerate
with respect to the background cosmological expansion. 
In sec. \ref{s-brems} the
quantum bremsstrahlung of gravitons at PBH collisions is discussed, which is quite similar to
the electromagnetic bremsstrahlung at Coulomb scattering of electrically charged particles.
Next, in sec. \ref{s-classical} we consider the classical emission of GW at accelerated motion 
of a pair of BHs in their mutual gravitational field. In sec.~\ref{s-energy-loss} we evaluate the
energy loss of PBHs due to their mutual interaction. It may be relevant to the estimation of the
probability of formation of PBH binaries. The gravitational radiation from PBH binaries in high
density clusters is discussed in  sec.~\ref{s-binaries}.  
In sec. \ref{s-evaporation} we calculate the present day energy density of gravitons produced 
at PBH evaporation. In sec. \ref{s-GW-background} we review some mechanisms of the 
production of stochastic background of GWs.  In sec. \ref{GW-detectors} the status 
of existing and planned detectors of GWs is discussed. In sec. \ref{conclusion} we conclude.
  
\section{Production and evolution of PBH in the early universe.} \label{s-BH-prod}

Formation of primordial black holes from the primordial density perturbations in the early
Universe was first considered by Zeldovich and
Novikov \cite{Novikov} 
and later by Hawking and Carr \cite{Hawking:1971ei, Carr:1974nx}. 
PBHs  would be formed 
when the density contrast, $\delta\rho /\rho$, at horizon was of the order of unity or,
in other words, when the Schwarzschild radius of the perturbation was of the order of the
horizon scale. If PBH was formed at the radiation dominated stage, when the cosmological
energy density was  $\rho(t) = 3 m_{Pl}^2/(32\pi t^2) $, and the horizon was 
$l_h = 2t$, the mass of PBH would be:
\begin{equation}
M(t)={ m_{Pl}^2t}\simeq 4\cdot 10^{38} \,\left(\frac{t}{\rm{sec}}\right)\,{\rm g}
\label{BH-mass}
\end{equation}
where $t$ is the time elapsed since Big Bang.

The fraction of the cosmological energy density of PBH produced by such mechanism depends upon the spectrum
of the primordial density perturbations. We denote this fraction $\Omega_p$ and take it as a free parameter of the 
model. The data on the large scale structure of the Universe and on the angular fluctuations of the cosmic
microwave background radiation (CMB) show that the spectrum of the primordial density fluctuations is almost 
flat Harrison-Zeldovich one. For such spectrum the probability of PBH production is quite low and
$\Omega_p \ll 1$.
However, the flatness of the spectrum is verified only for astronomically large scales, comparable with the
galactic ones. The form of the spectrum for masses below $10^{10}$ g is not known. Inflation predicts that
the spectrum remains flat for all the scales but there exist scenarios with strong deviation from flatness at small
scales. In particular, in ref.~\cite{Dolgov:1992pu, Dolgov:2008wu} a model of PBH formation has been proposed which leads to log-normal
mass spectrum of the produced PBH:
\be
\frac{dN}{dM} = C \exp \left[\frac{(M-M_0)^2}{M_1^2}\right],
\label{dN-dM}
\ee 
where $C$, $M_0$, and $M_1$ are some model dependent parameters. Quite
naturally the central value of PBH mass distribution may be in the desired range $M_0 < 10^{9}$ g.
In this model the value of $\Omega_p$ may be much larger than in the conventional model
based on the flat spectrum of the primordial fluctuations. We will not further speculate on the
value of $\Omega_p$ and on the form of the mass spectrum of PBH. In what follows we assume
for an order of magnitude estimate that the spectrum is well localized near some fixed 
mass value and that $\Omega_p$ is an arbitrary parameter. Different mechanisms of PBH production are reviewed e.g. in
ref.~\cite{Scenario1, Khlopov:2008qy}.

We assume that PBHs were produced in radiation dominated (RD) Universe, when the cosmological energy density 
was equal to
\be
\rho_{R} = \frac{3 m_{Pl}^2}{32 \pi t^2}\,.
\label{rho-RD}
\ee
If we neglect the PBH evaporation and possible coalescence, their number density would remain
constant in the co-moving volume, $n_{BH} (t) a^3(t) = const $. 
In what follows the instant decay approximation for evaporation is used. The cosmological evolution
of PBHs with more realistic account of their decay was studied in ref.~\cite{bcl-91}.

Since the black holes were non-relativistic at
production, their relative contribution to the cosmological energy density rose as the cosmological scale
factor, $a(t)$ :
\be
\Omega_{BH} (t) = \Omega_p \left(\frac{a(t)}{a_p}\right),
\label{Omega-of-t-3}
\ee
where $a_p$ is the value of the scale factor at the PBH production and at RD-stage
$a(t)/a_p = (t/t_p)^{1/2}$.  The moment $t_p$ of the black hole production is connected 
with the PBH mass through eq. (\ref{BH-mass}). Hence
\be
t_p = \frac{M}{ m_{Pl}^2}\,.
\label{t-p}
\ee
Thus if PBHs lived long enough, they would dominate the cosmological energy density 
and the Universe would become matter dominated at $t>t_{eq}$, where  
\be 
t_{eq} = \frac{M}{m_{Pl}^2 \Omega_p^2} = \frac{r_g}{2\Omega_p^2} \,,
\label{t-eq}
\ee
and $r_g = 2M/m_{Pl}^2$ is the gravitational (Schwartzschild) radius of a black hole.

In what follows we assume that all PBHs have the same mass $M$, but
the results can be simply generalized by integration over the PBH mass spectrum.

Evidently at RD stage the number density of PBHs drops as:
\be
n_{BH}(t) = n_p \left(\frac{a_p}{a(t)}\right)^3 = 
n_p \left(\frac{t_p}{t}\right)^{3/2}\,,
\label{n-of-t-RD}
\ee
while at MD stage
\be 
n_{BH}(t) =  n_p\left(\frac{t_p}{t_{eq}}\right)^{3/2}\,\left(\frac{t_{eq}}{t}\right)^2\,.
\label{n-of-t-MD}
\ee

Cosmological mass fraction of BH as a function of time behaves as
\be
\Omega_{BH} (t) = \frac{n_{BH}(t) M}{\rho_c} = \frac{16\pi}{3}\, r_g
t^2 n_{BH}(t) \,,
\label{Omega-of-t-2}
\ee
i.e. ${\Omega_{BH} \sim t^{1/2}}$ at RD stage. 
After the onset of the PBH dominance, 
$\Omega_{BH} $ approached unity and remained constant
till the PBH evaporation when 
${\Omega_{BH} }$ quickly dropped down to zero and the universe became
dominated by relativistic particles produced by PBH evaporation. All relics from the earlier RD stage would be diluted 
by the redshift factor $(t_{eq}/\tau_{BH})^{2/3}$. In particular the energy density of GWs produced at inflation would be diminished by this factor with respect to the standard predictions. Such dilution may cause problems with baryogenesis. However, these problems may be resolved if baryogenesis
took place at the process of PBH evaporation through the mechanism suggested by Zeldovich~\cite{zeld-BG} and quantitatively studied in ref.~\cite{dad-BG, dolgov1980}. Somewhat similar model of baryogenesis by heavy particle decay 
(e.g. by bosons of GUT) created at PBH evaporation was considered in ref.~\cite{barrow-BG, Barrow1981, Barrow1991, Baumann:2007yr}.

To survive till equilibration the PBHs should live long enough 
so that their evaporation time $t_{ev}$ would be larger than $t_{eq}$ or
$\tau_{BH}>t_{eq}-t_{p}$ which can be 
translated into the bound on the PBH mass:
\be
M >\left(\frac{N_{eff}}{3.2\cdot 10^4}\right)^{1/2}m_{Pl}
\,\left(\frac{1}{\Omega_p^2}-1\right)^{1/2}
\simeq  5.6\cdot 10^{-2}\,\left(\frac{N_{eff}}{100}\right)^{1/2}\frac{m_{Pl} }{\Omega_p}
\label{M-for-t-MD}
\ee
where $\Omega_p\ll 1$ and $M$ is mass of PBHs at production
\footnote{In fact in equation \eqref{M-for-t-MD} there must be the PBHs mass
 at the equilibrium time, $M(t_{eq})$.  Due to evaporation the  
PBH mass as a  function of time is given by
$M(t)=M(t_p)(1-t/\tau_{BH})^{1/3}$ and it is easy to see that for 
$\tau_{BH}>t_{eq}$ it gives $M=M(t_p)\simeq M(t_{eq})$, so hereafter
we refer to $M$ as the mass of PBH at production.}.
Both constraints (\ref{M-upper-bound}) and (\ref{M-for-t-MD}) would be satisfied if 
\be
\Omega_p > 0.7\cdot 10^{-14} \left(\frac{N_{eff}}{100}\right)^{1/6}.
\label{Omega-p-lower-limit}
\ee
For example, if $\Omega_p = 10^{-10}$, the black holes should be heavier than $1.2\cdot 10^{4}$ g. 

When the Universe  became dominated by non-relativistic 
PBHs, primordial density perturbations,
$\Delta=\delta\rho/\rho$,  should rise as the cosmological scale factor.
They could reach unity at cosmological time $t_1$ satisfying the condition:
\be 
\Delta_{in}\left(\frac{t_1} {t_{eq}}\right)^{2/3} \sim 1\, ,
\label{delta-rho}
\ee
where $\Delta_{in}$ is the initial magnitude of the primordial
density perturbations. To be more accurate, the evolution of density
perturbations depends upon the moment when they cross horizon, see
below, eq. (\ref{Delta-of-t1}). For the moment we neglect this
complication to make some simple estimates.

The initial density contrast is usually assumed to be of the order of
$ { \Delta_{in} \sim 10^{-5}-10^{-4}}$ which is not 
necessarily true at small scales and may be much larger, especially in the model
of ref.~\cite{Dolgov:1992pu,Dolgov:2008wu}. 

Evidently the BH life-time, ${\tau_{BH}}$, must be long
enough, so that the density fluctuations in BH matter would rise up
to the values of the order of unity. The condition $t_{ev}>t_1$ or equivalently $\tau_{BH} > t_1-t_p$
leads to the following restriction on the PBH mass:
\be 
M > M_{low}=\left(\frac{N_{eff}}{3.2\cdot 10^4}\right)^{1/2}\frac{m_{Pl}}{\Omega_p \Delta_{in}^{3/4}}
\simeq 1.2 \cdot 10^3 \,{\rm g}\,\left(\frac{10^{-6}}{\Omega_p}\right)
\left(\frac{10^{-4}} {\Delta_{in}}\right)^{3/4} \left(\frac{N_{eff}}{100} \right)^{1/2}\,.
\label{M-large-Delta}
\ee

We can see that eq. \eqref{M-large-Delta} puts a stronger lower
limit on PBHs mass than eq. \eqref{M-for-t-MD}. The limits are
comparable only if $\Delta_{in} \approx 1$.
Using eqs. \eqref{M-large-Delta} and \eqref{M-upper-bound} we get
a stronger than (\ref{Omega-p-lower-limit}) restriction on $\Omega_p$:
\begin{equation}
\Omega_p > 0.7\cdot 10^{-11}\left(\frac{10^{-4}}{\Delta_{in}}\right)^{3/4} \left(\frac{N_{eff}}{100}\right)^{1/6}.
\label{Omega-p-strongerlower-limit}
\end{equation}

After ${\Delta}$ reached unity, the rapid structure formation would take 
place and high density clusters of PBHs would be formed.
As we see in what follows, generation
of gravitational waves would be especially efficient from such high
density clusters of primordial black holes.

Let us assume that the spectrum of perturbations is the flat
Harrison-Zeldovich one and that a perturbation with some wave length
$\lambda$ crossed horizon at moment $t_{in}$. The mass inside horizon
at this moment was:
\be
M_b (t_{in} ) = m_{Pl}^2 t_{in}.
\label{M-b}
\ee
It is the mass of the would-be high density cluster of PBHs.
This initial time is supposed to be larger than $t_{eq}$ (\ref{t-eq}),
i.e. the horizon crossing took place already at MD-stage. For flat
spectrum of perturbations density contrast,
$\Delta = \delta\rho/\rho$, at horizon crossing
is the same for all wave lengths. After horizon crossing the
perturbations would continue to grow up as the scale factor,
$\Delta (t) = \Delta _{in} (t/t_{in})^{2/3}$. Such rise would
continue till moment $t_1(t_{in})$ such that:
\be
\Delta [t_1(t_{in})]= \Delta_{in} [t_1(t_{in}) /t_{in}]^{2/3} =1\,\,\, 
\rm{or}\,\,\, t_1 (t_{in} )=t_{in} \Delta_{in}^{-3/2}\,.
\label{Delta-of-t1}
\ee 

The radius of the PBH cluster rose almost as the cosmological scale factor
till $t=t_1(t_{in})$. After the density contrast has reached unity the cluster would
decouple from the common cosmological expansion. In other words, 
the cluster stopped to expand
together with the universe and, on the opposite, it would begin to shrink 
when gravity takes over the free streaming of PBHs. So the cluster size would drop down
and both $n_{BH}$ and $\rho_b$ would rise. The
density contrast would quickly rise from unity to $\Delta_b=\rho_b/\rho_c\gg 1$,
where $\rho_c$ and $\rho_b$ are respectively the average cosmological
energy density and the density of PBHs in the cluster (bunch). It looks
reasonable that  the density contrast of the evolved cluster could rise up
to ${\Delta = 10^{5}-10^{6}}$, as in the contemporary galaxies.
After the size of the cluster stabilized, 
the number density of PBH, $n_{BH}$, as well as 
their mass density, $\rho_{BH}$,
would be constant too. But the density contrast, $\Delta_b$ 
would continue to rise as $(t/t_1)^2$ because $\rho_c$ drops down as $1/t^2$. 
From time $t=t_1$ to $t=\tau_{BH}$ the density contrast would
additionally rise by the factor:
\be
\Delta (\tau_{BH})= \Delta (t_1) \left(\frac{\tau_{BH}}{t_1}\right)^2  = 
\Delta(t_i)\left(\frac{M}{M_{low}}\right)^4,
\label{Delta-of-tau}
\ee
where $t_1$ and $M_{low}$ are given by eqs.  (\ref{delta-rho} ) and
(\ref{M-large-Delta}) respectively.

The size of the high density clusters of PBH would be
\be
R_b = \Delta_b^{-1/3} t_1^{2/3} t_{in}^{1/3}\
\label{R-b}
\ee
and the average distance between the PBHs in the bunch can be estimated as:
\be
d_b = \left( M /M_b \right)^{1/3} R_b = \Delta_b^{-1/3} t_1^{2/3} r_g^{1/3}=
2^{-2/3} \Delta_b^{-1/3} \Delta_{in}^{-1} \Omega_p^{-4/3} r_g\,.
\label{d-b}
\ee
It does not depend upon $t_{in}$. Here eqs. (\ref{delta-rho}) and (\ref{t-eq}) have been used.

The virial velocity inside the cluster would be
\be 
v = \sqrt{\frac{2M_b}{m_{Pl}^2 R_b}} =2^{1/2} \Delta_b^{1/6} 
\Delta_{in}^{1/2} \approx 0.14
\left(\frac{\Delta_b}{10^6}\right)^{1/6}
\left(\frac{\Delta_{in}}{10^{-4}}\right)^{1/2}\,.
\label{v-b}
\ee
So PBHs in the cluster can be moderately relativistic.

Later, when $t=\tau_{BH}$, black holes would decay producing relativistic matter and the Universe
would return to the normal RD regime. However, the previous history of the earlier RD stage
would be forgotten.

For the future discussion it is convenient to introduce the average
distance between the PBHs at arbitrary time,
$d = n_{BH}^{-1/3}$, where $n_{BH} = \rho_{BH}/M$ is the number density of PBHs. Since
\be
\Omega_p = \frac{\rho_p}{\rho_c} = \frac{32\pi t_p^2 M n_p}{3
  m_{Pl}^2} = \frac{32\pi}{3}\,\left(\frac{t_p}{d_p}\right)^3 \,,
\label{Omega-p-of-d}
\ee
the average distance between PBHs at the production moment is equal to
\be
d_p = (4\pi/3)^{1/3} r_g \Omega_p^{-1/3}\,.
\label{d-p}
\ee
When the mutual gravitational attraction of PBH may be neglected, $d$ rises as cosmological scale
factor, $a(t)$.

Gravitational waves produced in the early universe will be hopefully registered in the present epoch. The sensitivity of GW detectors strongly depends upon the frequency of the signal. The frequency $f_*$ of GW produced at time $t_*$ during PBH evaporation, is redshifted down to the present day value, $f$, according to:
\be 
f = f_*\left[\frac{a(t_*)}{a_0}\right]=0.34\, f_*\, \frac{T_0}{T_{*}} \left[\frac{100}{g_S (T_*)}\right]^{1/3}\,,
\label{Omega-brems-today}
\ee
where $T_0= 2.725$ K \cite{Fixsen:2009ug} is the
temperature of the cosmic microwave background 
radiation at the present time,  $T_*\equiv T(t_{*})$ is the plasma
temperature at the moment of radiation of the gravitational waves, and
$g_S(T_*)$ is the number of species contributing to the entropy of the primeval plasma at temperature $T_*$.
It is convenient to express $T_0$ in frequency units, $T_0 = 2.7 \,{\rm  K} = 5.4\cdot 10^{10}$ Hz.
  
The temperature of the primeval plasma after the PBH evaporation
can be approximately found from:
\be 
\rho = \frac{m_{Pl}^2}{6\pi t^2} = \frac{\pi^2  g_*(T_*) T_*^4}{30}\,,
\label{rho-evap}
\ee
where ${g_* (T_*)\approx 10^2}$ is the contribution of
different particle species to the energy 
density at temperature ${T_*}$ and $t_1<t<t_{ev}$. 
For relativistic plasma $g_* (T) = g_S(T)$.
Since $t_{ev}=\tau_{BH}+t_p\simeq \tau_{BH}$, we obtain from
equation \eqref{rho-evap} at time $t_*=\tau_{BH}$:
\be 
T_*(\tau_{BH}) =  \left[\frac{30}{6\pi^3 g_S (T_*)}\right]^{{1}/{4}}
\left(\frac{N_{eff}}{3.2\cdot 10^4}\right)^{{1}/{2}}\,
\frac{m_{Pl}^{{5}/{2}}}{M^{{3}/{2}}}\,.
\label{T-tau-BH}
\ee

Substituting the numbers we find:
\be
T_*(\tau_{BH})
\approx 0.011 m_{Pl} \left[\frac{100}{g_S(T_*)}\right]^{1/4}
\left(\frac{N_{eff}}{100}\right)^{1/2}
\left(\frac{m_{Pl}}{M}\right)^{{3}/{2}}\,.
\label{tau-BH-num}
\ee
For comparison at the PBH production moment the temperature of the primeval plasma was:
\be 
T_p \approx 0.2 m_{Pl} \left(\frac{m_{Pl}}{M} \right)^{1/2} \,.
\label{T-p}
\ee
Using eqs. (\ref{Omega-brems-today}) and (\ref{tau-BH-num}), we find 
that the present day frequency of the GWs,  emitted at $T_*$ (\ref{T-tau-BH}) with frequency
$f_*$, would be equal to: 
\be 
f = 1.7\cdot 10^{12}\textrm{Hz} \left[\frac{100} {g_S(T_*)}\right]^{1/12} 
\left( \frac{100}{N_{eff}}\right)^{1/2}
\left(\frac{f_*}{m_{Pl}}\right)
\left(\frac{M}{m_{Pl}}\right)^{3/2}.
\label{omega-0}
\ee
If we take the maximum frequency of the emitted gravitons ${f_{max*}\approx r_g^{-1}= m_{Pl}^2/2M}$, the GW maximum frequency today would be:
\be
f_{max} \approx 8.6\cdot 10^{11}\textrm{Hz} \left(\frac{M}{m_{Pl}}\right)^{1/2}
= 5.8 \cdot 10^{16}\,{\rm Hz}\left(\frac{M}{10^5 {\rm g}}\right)^{1/2}\,.
\label{omega-today}
\ee

\section{Onset of GW radiation \label{s-onset} }

Once PBHs enter inside each other cosmological horizon\footnote{The cosmological horizon is the distance which PBHs started interacting with each other exchanging gravitons and should not be confused with the black hole event horizon.} they start to interact and
thus to radiate gravitational waves due 
to their mutual acceleration. The corresponding time moment $t_h$ is
determined  by the condition $2t_h=d(t_h)$ and remembering that 
it happened still at RD stage, we find
\begin{equation}
t_h = \frac{1}{2}\left(\frac{4\pi}{3}\right)^{2/3} r_g \Omega_p^{-2/3}.
\end{equation}
For ${t>t_h}$, the curvature effects can be neglected and the PBH
motion is completely determined by the Newtonian gravity:
\begin{equation}
\ddot{\mathbf{r}} = - \frac{M_{BH}}{m_{Pl}^2 r^2} \,\frac{\mathbf{r}}{r}
\label{ddot-r}
\end{equation}
with the initial conditions $r_i\equiv |\mathbf{r}_i|= d(t_i)$ and
$|\dot{\mathbf{r}_i}|=H(t_i)|\mathbf{r}_i|$, where $\mathbf{r}$ is the
position vector of PBHs. For ${t_{i} = t_h}$ their relative initial
velocity $\dot{ |\mathbf{r}_i}| = v_{i}=1$ and non-relativistic
approximation is invalid. To avoid that we should choose ${t_{i} >
  t_h}$ such that ${v_{i} \ll 1}$. The solution of the equation of
motion demonstrates that the effects of mutual attraction at this
stage and production of GW are weak.

After PBHs enter inside each other horizon and Newtonian gravity can
be applied, their acceleration toward each other becomes essential 
when their Hubble velocity drops below the capture velocity.
The corresponding time moment, $t_c$, when it happened, is 
determined from the condition:
\begin{equation}
\frac{1}{2}v^2(t_c) \equiv \frac{1}{2} [H(t_c) d(t_c)]^2 \lesssim \frac{M_{BH}}{m_{pl}^2d(t_c)}.
\label{velocity}
\end{equation} 
If it took place at the RD regime, 
the corresponding time moment would be equal to:
\begin{equation}
t_c = \frac{8\pi^2}{9}\,\frac{r_g}{\Omega_p^2},
\end{equation}
and the density parameter of PBHs at $t=t_c$ would be 
\begin{equation}
\Omega_{BH}(t_c) = \Omega_p \left(\frac{t_c}{t_p}\right)^{1/2} =
\frac{4\pi}{3} > 1\,.
\end{equation}
Thus at ${t=t_c}$ the universe is already matter dominated and we have
to use the non-relativistic expansion law, ${a \sim t^{2/3}}$, starting
from the moment $t=t_{eq}$ (\ref{t-eq}). Accordingly the  average distance
between BHs, when ${t>t_{eq}}$, grows as:
\be
d(t) = d_p
\left(\frac{t_{eq}}{t_p}\right)^{1/2}\,\left(\frac{t}{t_{eq}}\right)^{2/3}\,.
\label{d-of-t-MD}
\ee
Now we find that the condition that the Hubble
velocity, $v_H =(2/3t_c) d_c$
is smaller than the virial one, for average values, reads:
\be
\frac{4d_p^3}{9r_g t_p^{3/2} t_{eq}^{1/2}} <1\,.
\label{t2-nonrel}
\ee
One can see that this condition is never fulfilled. 
However, this negative result does not mean that the acceleration of
BHs and GW emission are suppressed, because of the mentioned above
effect of rising density perturbations.

\section{Bremsstrahlung of gravitons. \label{s-brems} }

PBH scattering in the early Universe should be accompanied by the graviton emission almost exactly as the scattering of charged 
particles is accompanied by the emission of photons. 
The cross-section of the graviton bremsstrahlung in particle collisions was calculated in 
ref.~\cite{Barker:1969jk} for the case of two spineless particles 
(here black holes) with masses $m$ and  $M$ under assumption that $m\ll M$. 
In non-relativistic approximation, $\mathbf{p}^2\ll m^2$, the differential cross section reads:
\begin{equation}
\mathrm{d}\sigma=\frac{64M^2m^2}{15m_{pl}^6}\frac{\mathrm{d}\xi}{\xi}
\left[ 5\sqrt{1-\xi} +\frac{3}{2} (2-\xi) \ln
  \frac{1+\sqrt{1-\xi}}{1-\sqrt{1-\xi}}\, \right]\,,
\label{sigma-brems}
\end{equation}
where $\xi$ is the ratio of the emitted graviton frequency,
$\omega=2\pi f$, to the kinetic energy of the incident black hole,
i.e. $\xi = 2m \omega / {\bf p}^2$.
We will use expression  (\ref{sigma-brems}) for an order of magnitude estimate assuming that it
is approximately valid 
for arbitrary $m$ and $M$, in particular, for $m\sim M$. 

The energy density of gravitational waves emitted at the time interval $t$ and $t+\mathrm{d}t$ 
in the  frequency range $\omega$ and $\omega+\mathrm{d}\omega$ is given by
\be
\frac{\mathrm{d}\rho_{GW}}{\mathrm{d}\omega} = 
 v_{rel} n_{BH}^2 \omega\left(\frac{\mathrm{d}\sigma}{\mathrm{d}\omega}\right)\mathrm{d}t\, ,
\label{d-dot-rho-domega}
\ee
where $n_{BH}$ is the number density of PBH and $v_{rel}$ is their
relative velocity.

The energy emitted in the frequency interval $\omega\in[0, \omega_{max}]$ 
per unit time is proportional to the integral
\begin{equation}
I(\omega_{max}) = \frac{{\bf p}^2}{2m}\varint_{0}^{\xi{max}} d\xi 
\left[ 5\sqrt{1-\xi} +\frac{3}{2} (2-\xi) \ln
  \frac{1+\sqrt{1-\xi}}{1-\sqrt{1-\xi}}\, \right]\,.
\label{integral}
\end{equation}

The maximum value of the frequency of the emitted gravitons should be smaller than 
either the kinetic energy of the colliding BHs, $E_{kin} = p^2/(2M)$ or the BH
inverse gravitational radius, $ 1/r_g =m_{Pl}^2/2 M$, depending on
which of the two is smaller. Their ratio is $E_{kin} r_g  = M^2 v^2 / m_{Pl}^2 $, so
for $M < m_{Pl} v^{-1} $ the maximum frequency would be  the PBH kinetic energy
and in this case $\xi_{max} =1$. It corresponds to the 
situation when PBH is nearly captured. It looses practically all its kinetic energy,
which goes to the graviton.
For PBHs in the high density clusters, when $v\sim 0.1$, the maximum frequency
would be $\omega_{max} \sim 1/r_g$ for all PBHs heavier than $10 m_{Pl}$. In this case
$\xi_{max}= (m_{Pl}/Mv)^2$.

The first, rather exotic case, when $M<m_{Pl}/v$
can be realized only if $\Omega_p \geq 0.01$, 
see eq. (\ref{M-for-t-MD}). If  $\xi_{max} =1$, then $\omega_{max} \sim {\bf p}^2/2m$
and the integral can be taken analytically:
\begin{equation}
I (\omega_{max}=p^2/2M) =\frac{25}{3}\frac{{\bf p}^2}{2m} = \frac{25}{3}\omega_{max}.
\label{I-of-omega-max}
\end{equation}
In this case the energy taken by GWs is of the order of the kinetic energy of PBH and correspondingly
$\Omega_{GW} \sim M n_{bh} v^2 /\rho_{BH} = v^2$.

Below we will consider more natural situation when $M > m_{Pl}\, v^{-1}$. Integral (\ref{integral}) 
in the limit of small $\xi_{max} $ is 
\be
I(\omega_{max}=1/r_g ) = \frac{p^2}{2M} \,\xi_{max}\,\left[8 + 3 \ln (4/\xi_{max})\right]
\label{I-of-omega-max-2}
\ee
This expression is accurate within 30\% up to $\xi_{max} =1$. So in what follows we will use 
this result as $I(\omega_{max}) \approx 25 \omega_{max}/3$, keeping in mind that normally
$\omega_{max} = 1/r_g \ll p^2/2M$.

The fraction of the cosmological
energy density of the emitted gravitational waves which has been produced during
time interval $t$ and $t+\mathrm{d}t$, which is smaller than or comparable to the cosmological
time $t_1\lesssim t\lesssim\ t_{ev}\simeq\tau_{BH}$,  can be obtained by the integration of 
equation  \eqref{d-dot-rho-domega}
over $\omega$ from $0$ to $\omega_{max}$ taking into account that the energy density of GWs
goes with the redshift as $(1+z)^{-4}$,  and the integration over cosmological
time, $t$, which is connected with the redshift by the relation\footnote{In this paper we consider flat space with 
curvature $k=0$ and neglect cosmological constant, $\Lambda=0$.}
\begin{equation}
\mathrm{d}t=-\frac{\mathrm{d}z}{H_*\,(1+z)\left[\Omega_{BH*}(1+z)^3+\Omega_{r*}(1+z)^4\right]^{1/2}}\,,
\label{redshift}
\end{equation}
where $H_*$, $\Omega_{BH*}$, and $\Omega_{r*}$ are respectively the Hubble parameter, the matter density 
parameter,  and the radiation density parameter evaluated at cosmological time $t_*=\tau_{BH}$, just
before the PBH decay. Recall that we use the instant decay approximation, so  
the Universe at $t = \tau_{BH}$  was still at MD stage. In this case all quantities such 
as $H_*$ and $\rho_c$ are taken at this stage:
$H_*=2/3t_*$,  $\rho_c=m_{Pl}^2/6\pi \tau_{BH}^2$, $\Omega_{BH*}=1$, and $\Omega_{r*}=0$.

We need to calculate the energy density of GWs at the moment of the PBH evaporation. The rate of GW
production is  given by eq. (\ref{d-dot-rho-domega}).  To take
into account the redshift of the energy density of the gravitational waves
we have to divide $d\rho_{GW}/d\omega$ by $(1+z)^4$, to substitute 
$\omega = (1+z)\omega_*$, where $\omega_*$ is the GW frequency at $t=\tau_{BH}$, and to express 
time through the redshift
as $\mathrm d t = (3/2)\tau_{BH}(1+z)^{-5/2} \mathrm d z$. As a result we obtain at $t_*=\tau_{BH}$:
\be 
\mathrm d\rho_{GW} (\tau_{BH}) = \frac{32 M^2 v_{rel}}{5 m_{Pl}^6} [\rho_{BH}^{(cluster)}]^2 
\tau_{BH} (1+z)^{-13/2} f[\omega_* (z+1)] \mathrm d[(1+z)\omega_*] \mathrm dz\,.
\label{d-rho-GW-brems}
\ee
Here $\rho_{BH}^{(cluster)}$ is the energy  density of the PBHs in the cluster (which is
denoted above as $\rho_b$). Note that $\rho_{BH}^{(cluster)} = const$ before the PBH decay. We parametrize 
this quantity as $\rho_{BH}^{(cluster)} = \rho_{BH}^{(c)} (\tau_{BH})\Delta(\tau_{BH})$, where 
$\rho_{BH}^{(c)} (\tau_{BH})=  m_{Pl}^2/(6\pi \tau^2_{BH})$ is the average cosmological energy density of PBH and 
$\Delta(\tau_{BH})$
is given by eq. (\ref{Delta-of-tau}), see also the discussion above this equation.
Function $f (\omega)$ is the function of $\xi = 2m\omega/p^2$ in the square brackets of eq. (\ref{sigma-brems}).

To find the cosmological energy fraction of GWs at $t=\tau_{BH}$ 
we need to integrate the expression above
over frequency, using eq.~(\ref{I-of-omega-max}),
and over redshift and to divide it by the total average cosmological energy
density $\rho_{BH}^{(c)} (\tau_{BH})=  m_{Pl}^2/(6\pi \tau^2_{BH})$. Since we have to average over the whole
cosmological volume, one factor $\Delta$ disappears and we remain with the first power of $\Delta$.
So the cosmological energy fraction of GWs would be:
\be
{\Omega_{GW} (\omega_{max}, \tau_{BH})}
\,\approx 16 Q\,  \left(\frac{v_{rel}}{0.1}\right)\,
\left(\frac{\Delta}{10^5}\right)\,
\left(\frac{N_{eff}}{100}\right)\, \left(\frac{\omega_{max}}{M}\right)\,.
\label{rho-brems}
\ee
Here coefficient $Q$ reflects the uncertainty in the 
cross-section due to the unaccounted for Sommerfeld enhancement~\cite{sommer, sakharov}.
Note that $\Delta$ may be considerably larger than $10^5$.

With ${v_{rel} = 0.1}$, ${\Delta = 10^5}$, $Q=100$, and $f_{max} = r_g^{-1}$ 
the fraction of the cosmological energy density of the GWs emitted by the 
bremsstrahlung of gravitons
from the PBHs collisions, when the Universe age was equal to the life-time of the
PBH, could reach:
\be
\Omega_{GW}(\tau_{BH}) \sim 3.8\cdot 10^{-17} \left(\frac{10^5\,\textrm{g}}{M}\right)^2\,.
\label{Omega-max}
\ee
It looks that for very light PBH, $M <  50 m_{Pl}$ , the fraction of
GW might exceed unity, which is evidently a senseless result. However,
one should remember the lower bound on the PBH mass (\ref{M-large-Delta}) and that
${ m_{Pl}/M < \Omega_p/20}$ and $m_{Pl}/ M < 10^{-7} (\Omega_p/ 10^{-6} ) $.

It may be interesting to calculate the contribution to $\Omega_{GW} (\tau_{BH}) $ from the earlier
period before the cluster formation. The mass density of PBHs at that stage was equal to the cosmological
energy density but since it was quite high and the effect is proportional to the density squared, the 
contribution from this period might be non-negligible. The result can be obtained from eq.~(\ref{d-rho-GW-brems}),
where $\rho_{BH}$ is taken equal to the average cosmological energy density. Since $\rho_{c}$ evolves with time
we need to insert into the integral over $\mathrm dz$ the factor $(1+z)^6$ where the redshift is taken from some initial
time, presumably $t_i=t_{eq}$, down to the moment of the cluster formation, $t_1$. So the energy density
of gravitational waves produced by bremsstrahlung from $t=t_{eq}$ (\ref{t-eq}) till $t=t_{1}$  (\ref{delta-rho})
would be:
\be 
\mathrm d\rho_{GW}^{(1)} = 
\frac{32 M^2 v_{rel}}{5 m_{Pl}^6} [\rho_{BH}^{(c)} (t_1)]^2 
t_{1} (1+z)^{-1/2} f[\omega_* (z+1)] \mathrm d[(1+z)\omega_*] \mathrm dz\,,
\label{d-rho-GW1-brems}
\ee
where $\rho_{BH}^{(c)} = m_{Pl}^2/(6\pi t_1^2)$ and $(1+z)$ runs from 1 up to $ (t_1/t_{eq})^{2/3}$. We have
introduced an upper index $(1)$ to indicate that this is the energy density of GWs generated before the cluster 
formation time  $t=t_1$. The integration over $z$ gives the enhancement factor 
$(1+z_{max})^{1/2} =  (t_1/t_{eq})^{1/3} $. According to 
eqs. (\ref{t-eq}) and (\ref{delta-rho}), this ratio is $\Delta_{in}^{-1/2} \sim 10^2$. Another enhancement
factor comes from a larger cosmological energy density $\rho^{(c)} (t_1) =\rho^{(c)} (\tau_{BH})(\tau_{BH}/t_1)^2$. 
The other factor $\rho_{BH}^{(c)} (t_1)$ disappears in the ratio $\Omega_{GW} = \rho_{GW}/\rho^{(c)}$.
On the other hand, $\Omega_{GW}$ is redshifted by $(\tau_{BH}/t_1)^{2/3}$.
Correspondingly 
\be
\frac{\Omega_{GW}^{(1)}(\tau_{BH})}{\Omega_{GW}(\tau_{BH})} = 
\frac{11 \Delta_{in}^{-1/2}} {\Delta (\tau_{BH})} \,\frac{v_{rel}^{(1)}}{v_{rel}} {\left(\frac{\tau_{BH}}{t_1}\right)^{1/3}}\,,
\label{Om1-over-Om}
\ee
where the coefficient 11 came from the ratio of the integrals over $z$ of eqs. (\ref{d-rho-GW-brems}) and 
(\ref{d-rho-GW1-brems}) and
\be
\left(\frac{\tau_{BH}}{t_1}\right)^{1/3} = \left(\frac{32170}{N_{eff} }\right)^{1/3} \Omega_p^{2/3} 
\left(\frac{M}{m_{Pl}} \right)^{2/3}\,.
\label{tau-over-t1}
\ee
The ratio of relative velocities of PBHs before and after the cluster formation, ${v_{rel}^{(1)}}/{v_{rel}}$,
is tiny, according to the estimates of sec.~\ref{s-onset}, and this introduces another strong suppression
factor to the production of GWs at an earlier stage.
In accordance with eq. (\ref{Delta-of-tau}) the density contrast rises as
$\Delta = \Delta(t_1) (\tau_{BH}/ t_1)^2$, where $\Delta(t_1)$ is supposed to be large, say, $10^4-10^5$
due to the fast rise of density perturbations at MD stage after they reached unity. Thus 
the generation of GWs in high density PBH clusters is much more efficient than at the earlier stage.

The density parameter of the gravitational waves at the present time 
is related to cosmological time $t_*$ as: 
\begin{equation}
\Omega_{GW}(t_0)=\Omega_{GW}(t_{*})\left(\frac{a(t_{*})}{a(t_0)}\right)^4\left(\frac{H_{*}}{H_0}\right)^2\,,
\label{Omega-GW-of-H} 
\end{equation}
where $H_0 =100 h_0 $~km/s/Mpc is the Hubble parameter and $h_0=0.74\pm 0.04$ \cite{Komatsu:2010fb, Nakamura:2010zzi}.

Using expression for redshift \eqref{Omega-brems-today} and taking the emission time $t_*=\tau_{BH}$ we obtain:
\begin{equation}
\Omega_{GW}(t_0)=1.67\times 10^{-5} h_0^{-2}\left(\frac{100}{g_S(T(\tau_{BH}))}\right)^{1/3}\Omega_{GW}(\tau_{BH})\,.
\label{generalexpression}
\end{equation}
Now using both equations $(\ref{Omega-max})$ and $(\ref{generalexpression})$ we find that
the total density parameter of gravitational waves 
integrated up to the maximum frequency is:
\be
h_0^{2}\Omega_{GW} (t_0) \approx 0.6\cdot 10^{-21}\,K
\left(\frac{10^5\,\textrm{g}}{M}\right)^{2}\,,
\label{Omega-brems-0}
\ee
where $K$ is the numerical coefficient:
\be
K = \left(\frac{v_{rel}}{0.1}\right)\,
\left(\frac{\Delta}{10^5}\right)\,
\left(\frac{N_{eff}}{100}\right)\,
\left(\frac{Q} {100}\right)\,
\left(\frac{100}{g_S (T(\tau_{BH}))}\right)^{1/3}\,.
\label{kappa}
\ee
Presumably $K$ is of order unity but since 
$\Delta$ may be much larger than one, see eq. (\ref{Delta-of-tau}), $K$ may also be large.

\section{GW from PBH scattering. Classical treatment. \label{s-classical}}

Classical radiation of gravitational waves by non-relativistic masses
is well described in quadrupole approximation, see e.g.  
books~\cite{Landau:1987gn, Misner:1974qy, Schutz:1985jx}. However,
as we have seen, in high density clusters of PBH, their relative velocity 
could be high, see eq.~(\ref{v-b}), and relativistic corrections may be 
non negligible. This problem was studied by Peters \cite{Peters:1970mx}, 
who considered emission of the GWs by two 
bodies with masses $M$ and $m$, where the former is supposed to be
heavy and at rest and the latter, lighter one, moves with velocity $v$.
For non-relativistic motion, when $v \ll 1$, and the minimal
distance between the bodies is larger than their gravitational radii,
the energy of gravitational waves emitted in a single scattering process is equal to:
\begin{equation}
\delta E_{GW}=\frac{37\pi}{15}\frac{M^2m^2v}{b^3m_{Pl}^6},\qquad v\ll 1\,,
\label{delta-E-nonrel}
\end{equation}
where $b$ is the impact parameter.

For the relativistic motion, $1-v^2 < 1$, the emitted energy is:
\begin{equation}
\delta E_{GW}=\frac{M^2m^2}{b^3m_{Pl}^6(1-v^2)^{3/2}}\,.
\label{delta-E-rel}
\end{equation}
The frequency of the emitted gravitational waves in this process is peaked
near $\omega \sim 2\pi/\delta t$, 
where $\delta t$ is the transition time which,  
for non-relativistic motion is $\delta t = b/v$ according to 
ref.~\cite{Peters:1970mx}, while for 
the relativistic one it is equal to $\delta t\sim b(1-v^2)^{1/2}$. 
For an order of magnitude estimate let us take $M\sim m$, then the
radiated energy, as a function of frequency, would be:
\begin{equation}
\delta E_{GW}(\omega)\approx\frac{M^4}{m_{Pl}^6} \,\omega^3\,.
\label{delta-E-of-omega} 
\end{equation}

This and the previous equations
are true for sufficiently large impact parameter, $b\gg r_g$ for which
the space-time between the scattered PBHs may be considered as flat and
their  gravitational mass defect can be neglected. The energy loss in
a single scattering event cannot be larger than 
\be
\delta E_{max} = \frac{p\,q}{M}\,,
\label{delta-E-max}
\ee
where $p = Mv_{rel}$ is the relative momentum of two scattered PBHs and $q$ is
the momentum transfer which by an order of magnitude is $q = 1/b$.  
Here and in what follows we use non-relativistic approximation.
So equations \eqref{delta-E-nonrel} and \eqref{delta-E-rel} can be true only 
for 
\be
b > b_{min}=\sqrt{\frac{37\pi}{15}} \,\frac{M^2}{m_{Pl}^3}\,.
\label{b-min}
\ee
For smaller impact parameters the radiation of gravitational waves would
be considerably stronger but the approximation used becomes invalid. For
the (near)  ``head-on'' collision of black holes a bound state of two BH (a binary)
or a larger black hole could be formed and the energy loss might be comparable
to the BH mass due to gravitational mass defect. However, we are
interested in gravitational waves at the low frequency part of
the spectrum, such that they could be registered by existing or
not-so-distant-future GW detectors. For such low frequency
gravitational waves the approximation used here is an adequate one.

The differential cross-section of the gravitational scattering of two PBHs in non-relativistic regime, 
$q^2\ll 2M^2$, can be taken as:
\be
\mathrm d\sigma = \frac{M^2}{m_{Pl}^2}\, \frac{\mathrm dq^2}{q^4}
=\frac{2M^2}{m_{Pl}^2} b \mathrm db\,  .
\label{d-sigma-dq2}
\ee
The differential energy density of GWs emitted at time and frequency intervals 
$[t,\,t+\mathrm{d}t]$ and  $[\omega,\,\, \omega+\mathrm d\omega]$ respectively 
can be calculated as follows. The rate of the energy emission by GWs is
\be
\mathrm d\dot\rho_{GW} = \mathrm d\sigma n^2_{BH} v_{rel} \delta E_{GW}\,,
\label{rho-dot-brems}
\ee
where we take for $\delta E$ non-relativistic expression (\ref{delta-E-nonrel}). We assume that the
impact parameter is related to the radiated frequency as $\omega = 2\pi v_{rel}/b$, as is discussed 
below eq. (\ref{delta-E-rel}). So $b\mathrm db = b^3\mathrm d\omega /(2\pi v_{rel})$. So we find:
\be 
\mathrm{d}\rho_{GW} = \frac{74 \pi v_{rel}} {15} \,\rho_{BH}^2\,
\frac{M^4}{m_{Pl}^8} \,\frac{\mathrm d\omega}{2\pi}\, \mathrm dt\,.
\label{d-dot-rho-GW}
\ee
The energy density parameter of GW at the moment of BH evaporation can be obtained integrating 
this expression over time and frequency. Thus we obtain:
 \be
\Omega_{GW} (\tau_{BH})= 2\cdot 10^{-10}\left(\frac{v_{rel}}{0.1}\right)^2 
\left(\frac{\Delta_b}{10^5}\right)\,
\left(\frac{N_{eff}}{100}\right)\,
\left(\frac{10^5\,\textrm{g}}{M}\right)\,.
\label{Omega-GW-of-t}
\ee
If we do not confine ourselves to the impact parameter bounded by
condition (\ref{b-min}) and allow for $b \sim r_g$, the energy
density of GWs at the moment of PBHs evaporation might be comparable to unity.
\begin{figure}[!htb]
\centering
\includegraphics[scale=.8]{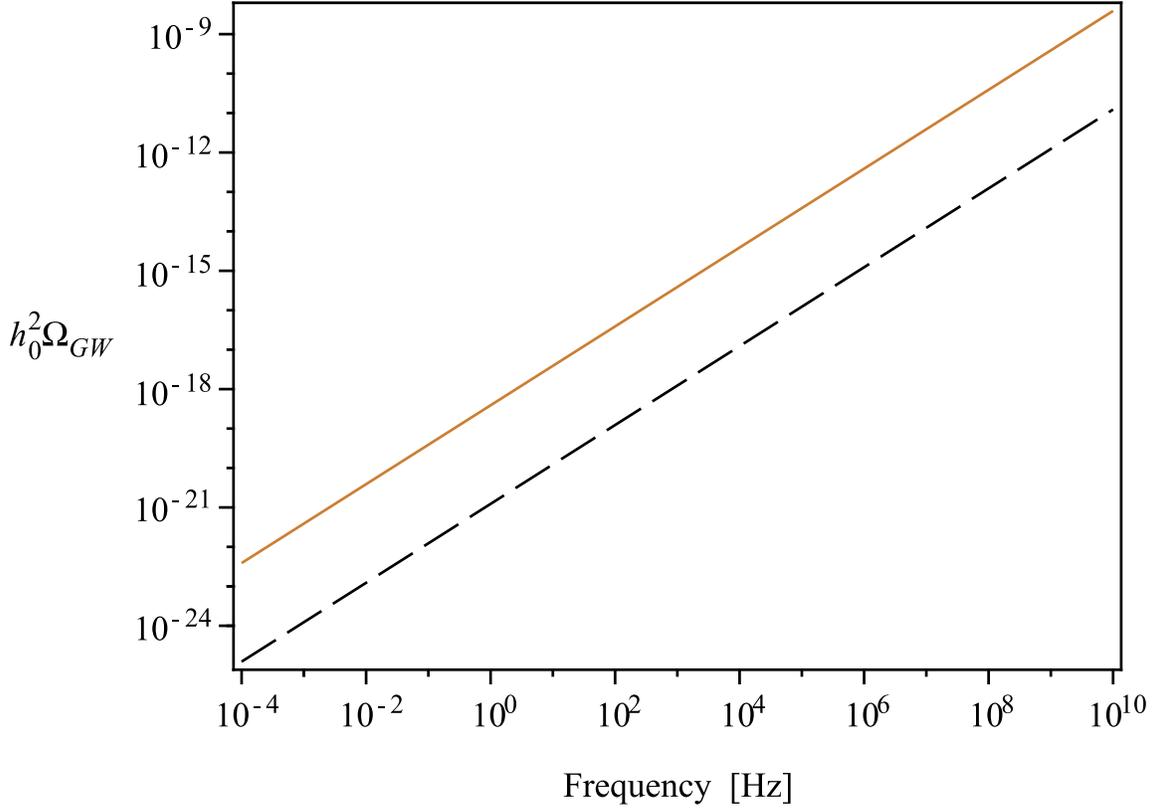}
\caption{Log-log plot of density parameter today, $h_0^{2}\Omega_{GW}$, as 
a function of expected frequency today in classical approximation for $N_{eff}\sim 100$,\, $g_S\sim 100$, 
$\Delta_b\sim 10^5$, and $v_{rel}\sim 0.1$ for different values of PBH mass $M\sim 1$ g (solid line) and $M\sim 10^5$ g (dashed line).}
\label{fig:classical}
\end{figure}

Let us now take into account the redshift of GWs emitted at different moments during the
the life-time of the high density clusters. The energy density of GWs emitted at some time
$t$ is redshifted to the moment of BH decay as $1/(z+1)^4$. The 
frequency of GW is redshifted as $\omega = (z+1)\omega_*$, where $\omega_*$ is the 
frequency of GWs at $t=\tau_{BH}$.  Integration over time or redshift is trivial and we find
from equation \eqref{d-dot-rho-GW} that the energy
density parameter of gravitational waves per logarithmic interval of frequency or the spectral density parameter, which is defined 
according to ref.~\cite{Thorne:1987} as:
\begin{equation}
\Omega_{GW}(f; t)\equiv \frac{1}{\rho_c}\,\frac{\mathrm{d}\rho_{GW}}{\mathrm{d}\ln f},
\label{spectraldensityparameter}
\end{equation}
at time $t=\tau_{BH}$ is equal to:
\be
\Omega_{GW} (f_*; \tau_{BH}) \approx 8.5 \,\left(\frac{v_{rel}}{0.1}\right) 
\left(\frac{\Delta_b}{10^5}\right)\,
\left(\frac{N_{eff}}{100}\right)\,
\left(\frac{M}{m_{Pl}^2}\right)\,f_*\,.
\label{Omega-GW-of-t-and-f}
\ee
Now using equations \eqref{omega-0} and \eqref{generalexpression}
we can calculate the relative energy density of GWs per logarithmic frequency at the present time:
\begin{equation}
h_0^{2}\Omega_{GW} (f; t_0)\approx 1.23\cdot 10^{-12}\alpha '\, \left(\frac{f}{\textrm{GHz}}\right)\,\left(\frac{10^5\,\textrm{g}}{M}\right)^{1/2}\,,
\label{Omega-GW-of-f-t_0}
\end{equation}
where $\alpha '$ is the coefficient at least of order of unity:
\begin{equation}
\alpha '= \left(\frac{v_{rel}}{0.1}\right)
\left(\frac{\Delta_b}{10^5}\right) 
\left(\frac{N_{eff}}{100}\right)^{3/2}
\left(\frac{100}{g_S(T(\tau_{BH}))}\right)^{1/4}\,.
\end{equation}
It may be much larger if $\Delta_b\gg 10^5$.

As we mentioned above, the classical approximation is valid if the impact parameter is bounded 
from below by equation \eqref{b-min}. Since the frequency of the radiated GWs is of the order of $v/b$, the 
maximum present day frequency of GWs, 
produced at cosmological time $t=\tau_{BH}$, 
for which the classical non-relativistic approximation is still valid, would be:
\begin{equation}
f_{max}\sim 9\cdot 10^5 \textrm{Hz}\left(\frac{v_{rel}}{0.1}\right)\left(\frac{100}{g_S(T(\tau_{BH}))}\right)^{1/12}
\left(\frac{100}{N_{eff}}\right)^{1/2}\left(\frac{10^5\,\textrm{g}}{M}\right)^{1/2}\,.
\end{equation}
For $M=10^5$ g the minimum impact parameter is $b_{min} \approx 10^{-13}$ cm. The frequency of the order 
of 1 Hz today corresponds to the impact parameter 6 orders of magnitude larger. If we demand that the impact parameter
should be smaller than the average distance  between PBHs in the clusters, then using equations (\ref{d-b}) and \eqref{b-min} we find that it can be true if the following condition is fulfilled:
\be
\Omega_p < 1.8 \cdot 10^{-6}\left(\frac{10^5 {\rm g}}{M}\right)^{3/4}\left(\frac{10^5}{\Delta_b}\right)^{1/4}\left(\frac{10^{-4}}{\Delta_{in}}\right)^{3/4}\,.
\label{Omega-p-M}
\ee

\section{Energy loss of PBHs \label{s-energy-loss}}

We calculate here the total energy loss of PBHs 
in the high density clusters, in order to understand how probable could be
the formation of the PBH binaries. 
First, let us estimate  the total energy loss of PBHs due to the
graviton bremsstrahlung. The loss of the kinetic energy per unit time
due to the  graviton emission is:
\be
-\left(\frac{\mathrm{d}E_{kin}}{\mathrm{d}t}\right)_{brem} = n_{BH} v_{rel} \varint_0^{\omega_{max}} \mathrm{d}\omega\,\omega
\left(\frac{\mathrm{d}\sigma}{\mathrm{d}\omega}\right)_{brem} \,,
\label{dot-E-kin}
\ee
where $\omega_{max}$ is defined in sec. \ref{s-brems}.
The total loss of kinetic energy of a single PBH 
during the time interval equal to the PBH life-time,
$\delta E_{kin} =-\dot E_{kin} \tau_{BH}$, normalized to the original kinetic energy
of the PBH can be estimated as
\be  
\frac{\delta E_{kin}}{E_{kin}} = 6\cdot 10^4 \kappa_2 \left(\frac{m_{Pl}}{M}\right)^2\,,
\label{delta-E-over-E}
\ee
where 
\be
\kappa_2 = \left(\frac{0.1}{v_{rel}}\right)\,
\left(\frac{\Delta_b}{10^5}\right)\,
\left(\frac{N_{eff}}{100}\right)\,
\left(\frac{Q} {10}\right)\, .
\label{kappa-2}
\ee
Clearly the energy loss is essential for very light PBHs which
could form dense clusters only if $\Omega_p$ is sufficiently high,
see eq.~(\ref{M-large-Delta}).

The energy loss due to classical GW emission might be somewhat more efficient.
According to the previous section the energy loss by a single PBH per
unit time is:
\be
 \Delta \dot E_{class} =
 n_{BH} v\varint_{b_{min}}^\infty \,db \left(\frac{\mathrm{d}\sigma}{\mathrm{d}b}\right)_{class}\delta E (b)\,,
\label{Delta-E}
\ee
where $\delta E (b)$ and $b_{min}$ are given respectively by eqs. (\ref{delta-E-nonrel})
and (\ref{b-min}).

Taking the integral over $b$ and time we find for the fractional energy loss of PBH due
to classical emission of the gravitational waves:
\be
\frac{\Delta E_{class}}{E_{kin} } = 0.9\cdot 10^3\, \frac{\Delta_b}{10^5}\,
\frac{N_{eff}} {100}\, \frac{m_{Pl}} {M}\,.
\label{Delta-E-class}
\ee
One should remember however that this energy loss comes from the PBHs scattering
with rather large impact parameter $b>b_{min}$. For smaller $b$, when the simple approximation
used in this work is inapplicable, the energy loss might be much larger.
Moreover, according to eqs. (\ref{t-eq}), (\ref{delta-rho}),  and (\ref{Delta-of-tau}) the density    
amplification factor $\Delta_b$ may be much larger than $10^{5}$:
\be
\Delta_b (\tau_{BH}) = 10^4\,\Delta (t_1)\, \Delta_{in}^3  \Omega_p^4 \left(\frac{100}{N_{eff}} \right)^2 
 \left(\frac{M}{m_{Pl}}\right)^4\,,
\label{Detla-amplf}
\ee
where we may expect e.g. that $\Delta (t_1) \sim10^5$, $\Delta_{in} \sim 10^{-4}$, and
$\Omega_{p} \sim 10^{-6}$. 

PBHs in the high density clouds could also loose their
energy by dynamical friction, see e.g. book~\cite{Binney:2008}. A
particle moving in the cloud of other particles would
transfer its energy to these particles due to their gravitational interaction. 
However, one should keep in mind that the case of
dynamical friction is essentially different from the energy loss due to
gravitational radiation. In the latter case the energy leaks out of the
system cooling it down, while dynamical friction does not change the
total energy of the cluster. Nevertheless a particular pair of black holes moving 
toward each other with acceleration may transmit their energy to the
rest of the system and became gravitationally captured forming a binary.

For an order of magnitude estimate we will use the
Chandrasekhar's formula which is valid for a heavy particle moving in
the gas of lighter particles having the Maxwellian velocity 
distribution with dispersion $\sigma$.  The  deceleration of 
a BH moving at velocity $v_{BH}$ with 
respect to the rest frame of the gas is given by
\begin{equation}
\label{eq:dynfric}
\frac{d}{dt}\vec{v}_{BH} = - 4 \pi \, G_N^2 \, M_{BH} \,
\rho_b  \, \ln \Lambda \, 
\frac{\vec{v}_{BH}}{v_{BH}^3} \, \left[ {\rm erf}(X) - 
\frac{2 X \exp(-X^2)}{\sqrt{\pi}} \right]\,,
\end{equation}
where $X \equiv v_{BH}/(\sqrt{2}\sigma)$, erf is the error function,
$\rho_b$ is the density of the background particles, and
$\ln\Lambda \approx \ln(M_*/M_{BH})$ is the Coulomb logarithm, which is 
defined as \cite{Binney:2008}: 
\begin{displaymath}
\ln\Lambda = \ln \frac{b_{max}  m_{Pl}^2\, \sigma^2}{M_{BH} + m} \, .
\end{displaymath}
Here $b_{max}$ is the maximum impact parameter, $\sigma^2$ is the 
mean square velocity of the gas and m is the mass of particles in the gas.
Numerical simulations show that $b_{max}$ can be assumed to be of the order
the radius of the cloud, $R_b$, which is given by equation \eqref{R-b}. Since 
$\sigma^2 \sim M_b/(m_{Pl}^2 R_b)$, a reasonable estimate of $\Lambda$ is 
$M_b/M_{BH}$. 

Equation \eqref{eq:dynfric} was solved in ref.~\cite{Bambi:2008uc} in two
limits $v>\sigma$ and $v<\sigma$. In both cases the characteristic 
dynamical friction time was of the order of:
\be
\tau_{DF} = \frac{\sigma^3 m_{Pl}^4}{4\pi\, M_{BH} \rho_b \ln \Lambda }
\approx \left(\frac{\sigma}{0.1}\right)^3 \left[\frac{25}{\ln (10^{-6}/\Omega_p)}\right]
\left(\frac{100}{N_{eff}}\right) \left(\frac{M}{1\,{\rm g}}\right)
\left(\frac{10^6}{\Delta}\right)\tau_{BH}\,.
\label{tau-DF}
\ee
For PBH masses below a few grams  dynamical friction would be an efficient mechanism 
of PBH cooling leading to frequent binary formation. Moreover,
dynamical friction could result in the collapse of small PBHs into much
larger BH with the mass of the order of $M_b$ (\ref{M-b}). This
process would be accompanied by a burst of GW emission.

\section{Gravitational waves from PBH binaries \label{s-binaries}}

Binary systems of PBH could be formed with 
non-negligible  probability in the high density clusters. 
As we have seen in the previous section, PBHs 
could loose their energy due to emission of gravitational waves and
due to dynamical friction~\cite{Binney:2008}. As a result  
they would be mutually captured. Determination of the capture probability is a complicated
task, which could probably be solved by numerical simulation. Since it is outside
of the scope of the present work, we simply assume that the mass  or number fraction of 
PBH binaries in the high density bunches of PBH is equal to $\epsilon$,
where  $\epsilon$ is a dimensionless parameter which is hopefully 
not too small  in comparison with unity.

Gravitationally bound systems of two massive bodies in circular orbit are known to emit gravitational 
waves with stationary rate and fixed frequency which is twice the 
rotation frequency of the orbit. In this approximation 
orbital frequency, $\omega_{orb}$, and orbit radius, $R$, are fixed. Luminosity of GW radiation from a single 
binary in the stationary approximation is well known, see e.g. book~\cite{Landau:1987gn}:
\be 
L_s \equiv \dot E = \frac{32 M_1^2 M_2^2 (M_1+M_2)}{5 R^5 m_{Pl}^8}= 
\frac{32}{5}\,m_{Pl}^2\left(\frac{M_c\,\omega_{orb}}{m_{Pl}^2}\right)^{10/3}\,,
\label{L-binary}
\ee
where $M_1$, $M_2$ are the masses of two bodies in the binary system and $M_c$ is the chirp mass which is defined as
\begin{equation}
M_c=\frac{(M_1\,M_2)^{3/5}}{(M_1+M_2)^{1/5}}
\end{equation}
and 
\be
\omega^2_{orb} =  \frac{M_1+M_2}{m_{Pl}^2 R^3}\,.
\label{omega-rot2}
\ee

In the case of elliptic orbit with large semi-axis $a$ and eccentricity $e$ the luminosity is 
somewhat larger (if $R=a$):
\be 
L_e = \frac{32 M_1^2 M_2^2 (M_1+M_2)}{5 a^5 m_{Pl}^8\, \left(1-e^2\right)^{7/2}}\,
\left(1+\frac{73 e^2}{24} +\frac{37e^4}{96} \right)\,.
\label{L-e--binary}
\ee
 
The emission of  GWs costs energy which is provided by the sum of the 
kinetic and potential energy of the system. To compensate the energy loss the radius of the binary 
system decreases and  the frequency rises making the stationary approximation invalid. 
As a result the system goes into the so called inspiral regime. 
Ultimately the two rotating bodies coalesce and produce a burst of gravitational waves. To reach this stage 
the characteristic time of the coalescence should be shorter than the life-time of the system. In our case it is 
the life-time of PBH with respect to the evaporation.

In the inspiral regime the initially circular orbit may remain approximately circular 
if radial velocity of the orbit, $\dot R$, is much smaller than the tangential velocity, $\omega_{orb}R$. 
This regime is called quasi-circular motion and is valid as long as (see e.g. book~\cite{Maggiore:1900zz}):
\begin{equation}
\dot\omega_{orb}\ll \omega_{orb}^2.
\label{quasi-circularcondition}
\end{equation} 
Equation \eqref{quasi-circularcondition} can be translated into the lower bound on the radius of the orbit:
\be
R\gg r_g^{(eff)} = \frac{M_1+M_2}{m_{Pl}^2}\,, 
\label{circular-2}
\ee
which is  the condition of the validity of the Newtonian approximation.
It was shown by Peters \cite{Peters:1963ux} that the orbits with initial $e_0=0$ would remain quasi-circular 
as far as condition \eqref{quasi-circularcondition} is fulfilled, 
while for the orbits with $e_0\neq 0$ the eccentricity rapidly approaches zero due to back reaction of the 
gravitational radiation. 

Most probably binaries are formed in elliptic orbits with high eccentricity. However in the calculation 
of the GW emission by binaries we assume for simplicity that all orbits are circular. 
The result would be a lower bound on GW emission, hopefully not too far from the real case.

In what follows we will consider both stationary and inspiral regimes 
since they both might be realized for different values of the
parameters. We will use the instant decay approximation, when the PBH mass
is supposed to be constant till $t=\tau_{BH}$ and then BH would
instantly disappear. The case of the realistic decrease of PBH mass will
be considered elsewhere.

The stationary orbit approximation would be valid if time of coalescence, $\tau_{co}$, would be much 
larger than the BH life-time, $\tau_{co}>\tau_{BH}$. The former can be found as follows (see e.g. 
book~\cite{Landau:1987gn}). According to the virial theorem the total (kinetic plus potential) energy 
of the system is ${\cal E} = -M_1 M_2 /(2R m_{Pl}^2)$. Since luminosity (\ref{L-binary})
is $L_s = -d{\cal E}/dt$, the radius varies with time  according to
\be
\dot R = - \frac{64 M_1 M_2 (M_1+M_2) } { 5 R^3 m_{Pl}^6}\,.
\label{dot-R}
\ee 
Correspondingly
\be
R(t) = R_0 \left( \frac{t_0 + \tau_{co} -t}{\tau_{co}} \right)^{1/4}\,,
\label{R-of-t}
\ee
where $R_0$ is the initial value of the radius, $t_0$ is the initial time, and the coalescence
time is given by:
\be
\tau_{co} = \frac{5 R_0^4\,m_{Pl}^6}{256 M_1 M_2 (M_1+M_2)}\,.
\label{t-coaelescence}
\ee

The condition $\tau_{co}>\tau_{BH}$ can be translated into the lower bound on $R$
(for $M_1=M_2$):
\be
R> R_{min} = 4.6\cdot 10^5\,\left(\frac{100}{N_{eff}}\right)^{1/4} 
\left(\frac{M}{10^5\,\textrm{g}}\right)^{1/2}\,r_g\,.
\label{R-min}
\ee
Keeping in mind that the frequency of GWs emitted at circular motion of the binary is twice 
the orbital frequency, $f_{s} = \omega_{orb}/\pi$ we find from 
equation \eqref{omega-rot2} that lower bound (\ref{R-min}) 
leads to the following upper bound on the GW frequency:
\be
f_s < \omega_{max}/\pi \approx 2\cdot 10^{24}\textrm{Hz}\,\left(\frac{N_{eff}}{100}\right)^{3/8}\left(\frac{10^5\,\textrm{g}}{M}\right)^{7/4}\,.
\label{omega-max}
\ee
On the other hand, the radius of the binary orbit should be smaller
than the average distance between PBHs in the cluster (\ref{d-b}) and probably quite close to it.
Using eqs. (\ref{d-b}) and (\ref{R-min}) we find: 
\be
\frac{R_{min}} {d_b} = 1.3\cdot 10^{-5} \left(\frac{\Delta_b}{10^5}\right)^{1/3}\,
\left(\frac{\Delta_{in}}{10^{-4}}\right)\, \left(\frac{\Omega_p}{10^{-6}}\right)^{4/3}\,
\left(\frac{M}{10^5 g}\right)^{1/2}\,.
\label{Rmin-over-db}
\ee
So it seems natural that $R_{min} \ll d_b$ and the PBH binaries should be mostly in
the quasi-stationary regime. $R_{min}$ would be equal to $d_b$ roughly speaking for quite 
large mass fraction of the produced PBHs, $\Omega_p > 10^{-3}$.

The condition $R_{min} = d_b$ gives a lower bound on orbital frequency, $\omega_{orb}$:
\begin{equation}
\omega_{orb}> \omega_{min}\approx 9.4\cdot 10^{17}\, \textrm{sec}^{-1}\,\left(\frac{\Delta_b}{10^5}\right)^{1/2}\left(\frac{\Delta_{in}}{10^{-4}}\right)^{3/2}
\left(\frac{\Omega_p}{10^{-6}}\right)^2\,\left(\frac{10^5\,\textrm{g}}{M}\right)\,.
\label{omega-min}
\end{equation}

During the inspiral phase, for which $\tau_{co}<\tau_{BH}$, we expect that binaries emit GWs in the frequency range:
\begin{equation}
2\cdot 10^{24}\textrm{Hz}\,\left(\frac{N_{eff}}{100}\right)^{3/8}\left(\frac{10^5\,\textrm{g}}{M}\right)^{7/4}<
f < 0.6\cdot 10^{33}\textrm{Hz}\,\left(\frac{10^5\,\textrm{g}}{M}\right)\,.
\label{f-range-insp}
\end{equation}
The upper bound corresponds to $\omega \sim 1/r_g$.

The frequency spectrum of the gravitational waves in inspiral but quasi-circular motion can be found 
in the adiabatic approximation as follows. Since the gravitational
waves are emitted in a narrow band near twice the orbital frequency,
the spectrum of the luminosity (\ref{L-binary}) can be approximated as:
\be
\mathrm d \dot E = \frac{32 M_1^2 M_2^2 (M_1+M_2)} {5 R^5 (t) m_{Pl}^8} \,
\delta \left( \omega - 2\omega_{orb} (R) \right) \mathrm d\omega
\label{d-dot-E}
\ee
To find the energy spectrum we have to integrate this expression over
time from initial time, $t_{min}=t_0$, to maximum time 
$t_{max} = min [\tau_{BH}+ t_p, \tau_{co}+t_0]$, where $t_0$ and $t_p$
are respectively the time of the binary formation (it may be different
for different binaries but here we neglect this possible spread) and the time of PBH formation (it is different
for PBH with different masses). Note that the coalescence time, $\tau_{co}$
is also different for binaries with different initial radius $R_0$. 

Using eqs. (\ref{omega-rot2}) and (\ref{dot-R}) and the expression 
$\mathrm dt = (\mathrm dR/\mathrm dt)^{-1} (\mathrm dR/\mathrm d\omega_{orb}) \mathrm d\omega_{orb}$, we find:
\be
\frac{\mathrm d E}{\mathrm d\ln \omega} = \frac{2^{1/3}\omega^{2/3}}{3}\frac{M_1M_2} {m_{Pl}^{4/3}(M_1+M_2)^{1/3}}
\label{dE-dlog-bin}
\ee
in agreement with refs.~\cite{Maggiore:1900zz}, \cite{Phinney:2001di}. This expression is
valid for the frequencies in the interval determined by eq. (\ref{omega-rot2}) with $R_{max}=R_0$
and $R_{min}=R(t_{max})$. 
 
In expression (\ref{dE-dlog-bin}) we have not taken in account the redshift which is different
for different frequencies and thus this leads to spectrum distortion. According to 
eqs. (\ref{omega-rot2}) and (\ref{R-of-t}) frequency $\omega$ is emitted at the 
time moment:
\be
t(\omega) = t_0 +\tau_{co} \left[ 1 - \left( \frac{ \omega_{min}} {\omega} \right)^{8/3} \right]\,,
\label{t-of-omega}
\ee
where 
\be
\omega_{min} = 2 \left(\frac{M_1+M_2}{m_{Pl}^2} \right)^{1/2} \, R_0^{-3/2}
\label{omega-min-2}
\ee
is the minimal frequency emitted at initial moment $t=t_0$.
To the moment of the PBH evaporation the frequency of the GWs emitted at $t=t(\omega)$
is redshifted by the frequency dependent factor:
\be
\omega_* = \frac{\omega}{ 1+ z(\omega) } = \left[\frac{t (\omega)} {t_p+\tau_{BH} }\right]^{2/3} \omega,
\label{omega-star}
\ee
where $\omega_*$ is the frequency of GWs at $t= t_p + \tau_{BH}$. This equation
implicitly determines $\omega$ as a function of $\omega_*$.

The spectrum of the gravitational waves at PBH evaporation can be obtained from eq. (\ref{dE-dlog-bin}) 
dividing it by $(1+z)$ (the redshift of the graviton energy, $E$)
and with substitution $\omega = (z+1) \omega_*$. Correspondingly
\be
\mathrm d\omega = \frac{z+1}{1-\omega_* (\mathrm dz/\mathrm d\omega) }\,\mathrm d\omega_*
\label{d-omega}
\ee
As a result we find:
\be
\frac{\mathrm d E_*}{\mathrm d\ln \omega_*} = \frac{2^{1/3}\omega_*^{2/3}}{3}\frac{M_1M_2} {m_{Pl}^{4/3}(M_1+M_2)^{1/3}}\,
\frac{\left[1-\omega_* (\mathrm dz/\mathrm d\omega)\right]^{-1}}{(1 +z)^{1/3}}\,.
\label{dE*-dlog}
\ee
Here $z(\omega)$ should be taken as a function of $\omega_*$ according to eq. (\ref{omega-star}) and
$\omega_*$ varies between $\omega_{min} $ and $\omega_{max}$ divided by the corresponding red-shift
factor. In particular, $\omega_{* (min)} = \omega_{min} [t_0/(t_p+\tau_{BH})]^{2/3}$.
Note that $R_0$ enters explicitly into eq. (\ref{dE*-dlog}), while in eq. (\ref{dE-dlog-bin}) it enters only through 
the limits in which  $\omega$ varies. Because of that the frequency spectrum depends upon the distribution
of binaries over their initial radius, $R_0$. As is shown below, it is especially profound in the case of long
coalescence time when the frequency spectrum of a single binary with fixed $R$ is close to delta-function.

In the stationary approximation, when 
the change of the orbit radius can be neglected, we expect that a single binary emits 
GWs in a narrow band of frequencies close to twice the orbital frequency. However the
distribution of binaries over their initial
radius, $ \mathrm dn_{BIN} = F(R_0) \mathrm dR_0 $  spreads up the spectrum. Here
$\mathrm dn_{BIN}$ is the number density of binaries with the radius in the interval $[R_0, R_0+\mathrm dR_0]$. 
Since in this approximation the radius is approximately constant we do not distinguish between $R$ and $R_0$. 
The cosmological energy density of the  gravitational waves 
emitted per unit time is equal to:
\be
\mathrm d \dot\rho_{GW}^{(stat)} =\frac{2F(R)\,R}{3} \frac{n_{BH}^c}{n_{BH}^b}\,\frac{\mathrm d\omega}{\omega}\, L_s\,, 
\label{d-dot-rho1}
\ee
where $n_{BH}^b$ is the number density of PBH in the high density bunch (cluster),
$n_{BH}^c$ is the average cosmological number density of PBH,
$R = R(\omega_{orb})$ according to eq. (\ref{omega-rot2}), and we used the relation
$\mathrm dR = -2(R/3)\,(\mathrm d\omega/\omega)$. Distribution, $F(R)$, is normalized as:
\be
\int \mathrm dR F(R) = n_{BIN} = \epsilon n_{BH}^b\,.
\label{int-F-of-R}
\ee
We assume for simplicity that $F(R) $ does not depend upon $R$ in some
interval $[R_1,\,R_2]$ and vanishes outside it. 
So $F(R) = \epsilon n_{BH}^b/(R_1-R_2)$.

A more realistic fit to the PBH distribution over radius could be a Gaussian one:
\be
F(R) = \frac{ 1 } {\sqrt{2\pi} \,\sigma} \epsilon n_{BH} 
\exp\left[ - { \left( R- \langle R \rangle \right)^2 }/ {2 \sigma^2}\right]\,,
\label{F-of-R-gauss}
\ee
where $\sigma$ is the mean-square deviation of $R$ from the average value $\langle R \rangle$.

The small factor $n_{BH}^c/n_{BH}^b $ enters eq.~(\ref{d-dot-rho1})
because we are interested in 
the cosmological energy density of GWs  averaged over the whole universe volume.
The cosmological number density of PBH is expressed through their energy density
as $n_{BH} = \rho_{BH}/M = \rho_c (t)/M$.
The number density of binaries in the cluster is parametrized according to:
\be
n_{BIN}(t) = \epsilon (t)\, n^b_{BH}(t) =\epsilon (t)\, \rho_c(t)
\Delta (t) /M \,,
\label{n-BIN}
\ee
where, we remind, $\rho_c (t)$ is the total cosmological energy density and 
$\Delta (t) = \rho_b /\rho_c \gg 1$ is the density
contrast of the cluster. 
The time dependence of $n_{BH}^b$  
disappears when the cluster reaches the stationary state,
see discussion in Sec.~\ref{s-BH-prod}, and $\Delta (t)$ evolves according to
equation~(\ref{Delta-of-tau}). When the stationary orbit approximation is valid,
$\epsilon $ remains constant.

Collecting all the factors and integrating eq. (\ref{d-dot-rho1}) over time with an account of the
frequency redshift, $\omega = \omega_* (1+z)$ and the total redshift of the energy
density of GWs, $\rho_{GW}(t_*)=\rho_{GW}(t)/(1+z)^4$, we find:
\be
\mathrm d \rho_{GW}^{(stat)}(\omega_*; \tau_{BH})=
 \frac{2^{7/3}}{5}\left[\frac{n^c_{BH}(\tau_{BH})}{n^b_{BH}}\right]\frac{(M_1^2 M_2^2)(\tau_{BH}+t_p)}{(M_1+M_2)^{1/3} m_{Pl}^{16/3}}\,F(R) \,\omega_*^{5/3} \mathrm d\omega_*\, 
\varint_{x_{min}}^{1} x^{11/6}\,\mathrm dx\,,
\label{d-rho-stat}
\ee
where $x=a(t)/a(t_*)=1/(1+z)$, $x_{min}=a(t_0)/a(t_*)$, $t_0$ is the time moment of binary formation and we make use of equation \eqref{redshift}.
Dividing this result by the critical energy density just before PBHs complete evaporation,
$n_{BH}(\tau_{BH})\approx \rho_c(\tau_{BH})/M$, we find the cosmological fraction of the energy density of GWs at $t=\tau_{BH}$ per logarithmic interval of frequency $f = \omega/(2\pi)$ (below we assume that all BHs have equal masses, $M$):
\be
\Omega_{GW}^{(stat) }(f_*; \tau_{BH})  = \frac{3\cdot 2^{17/3}}{85}\,
\frac{\epsilon\cdot(t_p+\tau_{BH})}{R_1-R_2}\,
\left(\frac{\pi f_* M} {m_{Pl}^2} \right)^{8/3}\,[1-x_{min}^{17/6}]
\label{dOmega-dlogf*}
\ee
where for the sake of a simple estimate we assumed that $F(R) = const$. We assume also that all the binaries are formed at the same time, $t_0\ll \tau_{BH}$ and so $x_{min}\ll 1$.
Note that the frequency of GWs coming from the binaries with radii between $R_1$ and $R_2$ is confined according
to eq. (\ref{omega-rot2}).

To make an order of magnitude estimate of the fraction of the energy density of GWs at the moment of PBH evaporation we take $(R_1 -R_2) \sim R_1 \sim R(\omega)$, where $R(\omega)$ is determined
by equation (\ref{omega-rot2}) and take into account that the stationary approximation is valid if the radii of the binaries are bounded from below by eq.~(\ref{R-min}). Hence, if the stationary regime is realized, the spectral density parameter today would be:
\be
h_0^{2}\Omega_{GW}^{(stat)} (f; t_0) \approx 10^{-8} \epsilon 
\left[\frac{N_{eff}}{100}\right]^{2/3}\left[\frac{100}{g_S(T(\tau_{BH}))}\right]^{1/18} \left[\frac{M}{10^5\,{\rm g}}\right]^{1/3}\,\left[\frac{f}{{\rm {GHz}}}\right]^{10/3}
\label{Omega-upper-limit}
\ee
\begin{figure}[!htb]
\centering
\includegraphics[scale=.8]{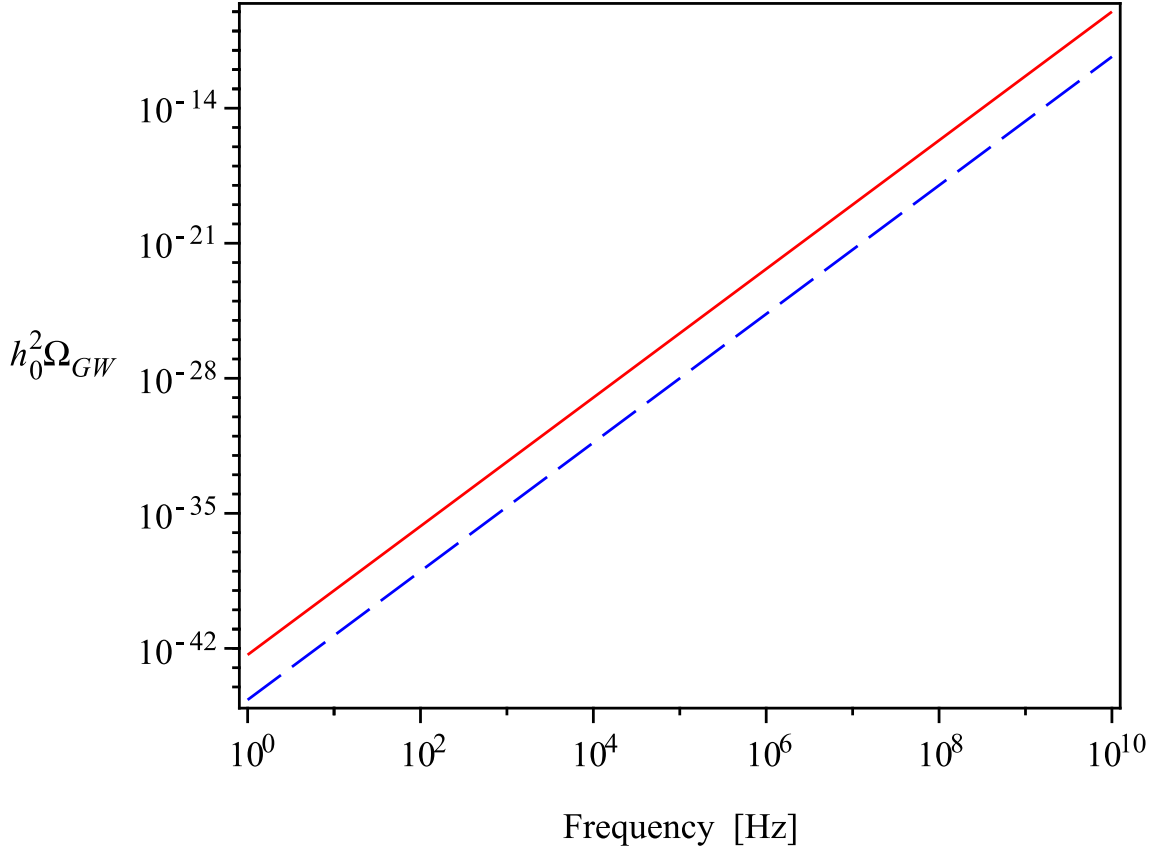}
\caption{Log-log plot of density parameter today $h_0^{2}\Omega_{GW}$ as a function of expected frequency today for 
PBHs binaries in the stationary approximation for $\beta\sim 1$, $\epsilon\sim 10^{-5}$, $N_{eff}\sim 100$, $g_S\sim 100$, PBH mass $M\sim 10^7$ g (solid line) and $M\sim 1$ g (dashed line).}
\label{fig:binaries}
\end{figure}

The expected range of the present day frequencies of the GWs from the binaries in the stationary approximation 
is given by eqs. \eqref{omega-min} and \eqref{omega-max}. 
The emitted frequency is determined by the binary radius, so a single binary emits GWs  with a very narrow spectrum.
However, the distribution of binaries over their radius could lead to a significant spread of the spectrum. 
In principle the frequencies emitted may have any value in the specified above range. 
The minimal present day frequency of such GWs today can be found by plugging eq. \eqref{omega-min} 
into eq. \eqref{omega-0}:
\begin{equation}
f_{}\geq 4.3\, \textrm{Hz}\,\beta\,\left(\frac{10^5\,\textrm{g}}{M}\right)^{1/2}\,,
\end{equation}
where $\beta$ is given by
\begin{equation}
\beta=\left(\frac{\Delta_b}{10^5}\right)^{1/2}\left(\frac{\Delta_{in}}{10^{-4}}\right)^{3/2}
\left(\frac{\Omega_p}{10^{-6}}\right)^2\left(\frac{100}{g_S(T(\tau_{BH}))}\right)^{1/12}\left(\frac{100}{N_{eff}}\right)^{1/2}\,.
\end{equation}
For binaries formed with $R>R_{min}$, see equations \eqref{omega-0}, \eqref{R-min} and \eqref{omega-max}, 
the frequency of emitted GWs today is bounded from above by:
\begin{equation}
f_{}\leq 5.7 \cdot 10^7\,\textrm{Hz}\,\left(\frac{100}{g_S(T(\tau_{BH}))}\right)^{1/12}\left(\frac{100}{N_{eff}}\right)^{1/8}\left(\frac{10^5\,\textrm{g}}{M}\right)^{1/4}\,.
\end{equation}

Let us estimate now the energy density of GWs in the inspiral case, when $\tau_{co}< \tau_{BH}$ and 
the GW emission from a single binary proceeds in a wide range of frequencies due to shrinking of the binary radius.
The radiation frequency spans from $f_{s, min}$, which is the GW frequency at the initial PBH separation, 
to $f_{s, max}$ which corresponds to GWs emitted at $R\sim r_g$.
The energy spectrum of GWs is given by eq.   (\ref{dE-dlog-bin}) where, in what follows, we change
to cyclic frequency, $f =\omega/2\pi$. 

After the cluster evolution was over, the number density of PBHs in high density clusters remained approximately 
constant till the PBH evaporation, but in the inspiral phase the fraction of binaries, $\epsilon (t)$, decreased due to
their coalescence. So the tail of the distribution function at small initial $R_0$ is eaten up, and the average value 
of $R$ drops down. In distribution function, $F(R_0)$, we have to substitute instead $R_0 $ its expression through
$R$ and time according to 
\be
R_0 \rar \left[ R^4 + \left(\frac{256 M_1M_2 (M_1+M_2)} {5 m_{Pl}^6}\right)\,(t-t_0)\right]^{1/4}
\label{R0-of-R}
\ee
with the corresponding change of $R^3_0 \mathrm dR_0  \rar  R^3 \mathrm dR$. 

To calculate the cosmological energy fraction of GWs at the PBH evaporation moment we can proceed 
along the same lines as we have done deriving eq.  (\ref{dE*-dlog})  introducing additional factor
$F(R_0) \mathrm dR_0$ which depends upon time according to eq. (\ref{R0-of-R}). However, at the level
of calculations in the present model with many unknown parameters it can be sufficient to neglect 
such subtleties and to use a simplified estimate:
\begin{equation}
\frac{\mathrm{d}\rho_{GW}}{\mathrm{d}(\log f_s)}=\epsilon_{co} n_{BH}^c(t)\frac{\mathrm{d}E_{GW}}{\mathrm{d}(\log f_s)}\,,
\end{equation}
where $\epsilon_{co}$ is the fraction of binaries  with 
coalescence time shorter or equal to PBH life-time. For an estimate by an order of magnitude we assume
also that the number of binaries is independent on the redshift. To some extend the decrease of the binary
number may be compensated by their continuous formation. We neglect possible difference of binary 
masses and take $M_1 = M_2$. We approximately take the redshift into account from the moment
of the coalescence to the PBH decay, $(z_{co}+1) \approx (\tau_{BH}/\tau_{co})^{2/3}$. This corresponds
to the assumption that
the binaries radiated all GWs only at the moment of $\tau_{co}$. So the $f_* = f (1+z_{co})$. 
Thus we obtain as an order of magnitude estimate:
 \be 
\Omega_{GW}(f_*,\tau_{BH}) = \frac{\epsilon_{co} }{3}
\left( \frac{\pi f_* M} {m_{Pl}^2 } \right)^{2/3}\, (z_{co}+1)^{-1/3}\,.
\label{Omega-GW-insp}
\ee

Using equations \eqref{omega-0} and \eqref{generalexpression},
we find that the energy density parameter of gravitational waves
today is equal to:
\begin{equation}
 h_0^{2}\Omega_{GW}(f)\approx 5 \cdot 10^{-9}\epsilon_{co}
\left(\frac{100}{g_S(T(\tau_{BH}))}\right)^{5/18}\left(\frac{N_{eff}}{100}\right)^{1/3}  
\left(\frac{f}{10^{12}\textrm{Hz}}\right)^{2/3}\left(\frac{10^5\,\textrm{g}}{M}\right)^{1/3}\,,
\end{equation}
where we neglected possibly weak redshift dilution of GWs by the factor $(\tau_{co}/\tau_{BH})^{2/9}$. 
   
If the system goes to the inspiral phase, then according to equation (\ref{f-range-insp}) we would expect today  
a continuous spectrum in the range from $f_{min}\sim 0.9\cdot 10^7 \textrm{Hz} $ to $f_{max}\sim 3\cdot 10^{14}\,\textrm{Hz}$. 
However if we take into account the redshift of the early formed binaries from the moment of their formation to the PBH decay, the lower value of the frequency may move to about 1 Hz.

\section{Gravitons from PBH evaporation \label{s-evaporation}}

In the previous sections we have considered only gravitational waves emitted through mutual acceleration of PBHs in the high density clusters. 
On the other hand PBHs could directly produce gravitons by evaporation. This process in connection with creation of cosmological background of relic GWs
was considered in ref.~\cite{Dolgov:2000ht} and later in ref.~\cite{Anantua:2008am}. In the last reference a possible clumping 
of PBHs at the matter dominated stage was also considered. Though such clumping does not influence the probability of the GW emission by PBHs, it may change the mass spectrum of PBHs due to their merging.

The PBHs reduce their mass according to the equation:
\begin{equation}
M(t)=M_0\left(1-\frac{t-t_p}{\tau_{BH}}\right)^{1/3}\,,
\label{reducingmass}
\end{equation}
where $M_0$ is the initial mass of an evaporating BH and $t_p$ is the time of BH production after Big Bang. 
Equation \eqref{reducingmass} shows that the BH mass can be approximately considered as constant till 
the moment of the evaporation and may be 
approximated as  $\theta (t-t_p-\tau_{BH})$. Due to evaporation a BH emits all kind of particles with masses $m<T_{BH}$
and, in particular, gravitons. 
The total energy emitted by BH per unit time and frequency $\omega$ (energy) of the emitted particles, 
is approximately given by the equation (see, e. g. book \cite{Frolov:1998wf}):
\begin{equation}
\left(\frac{\mathrm{d}E}{\mathrm{d}t\mathrm{d}\omega}\right)=
\frac{2N_{eff}}{\pi}\,\frac{M^2 }{m_{Pl}^4}\frac{\omega^3}{e^{\omega/T_{BH}}-1}\,,
\end{equation}
where $T$ is the BH temperature \eqref{T-BH}. 
Due to the impact of the gravitational field of BH on the propagation of the evaporated particles, their spectrum 
is distorted \cite{DonPage:1976} by the so called grey factor $g(\omega)$, but we disregard it in what follows. 

Let us now estimate the amount of the gravitational radiation from the graviton evaporation. 
After their production PBHs started to emit thermal gravitons independently on the PBH clustering. 
Hence the thermal graviton emission depends only on PBH number density, $n_{BH}$. 
The energy density of gravitons in logarithmic frequency band emitted in the time 
interval $t$ and $t+ \mathrm d t$ is
\begin{equation}
\frac{\mathrm d \rho_{GW}(\omega; t)}{\mathrm d \omega}=
10^{-2}n_{BH}(t)\,\left(\frac{\mathrm{d}E}{\mathrm{d}t\mathrm\,{d}\omega}\right)\mathrm d t\,,
\label{d-rho-evap}
\end{equation}
where factor $10^{-2}$ takes into account that  about one percent of the emitted energy goes into gravitons.
The density parameter of GWs per logarithmic frequency interval at cosmological time $t_* =\tau_{BH}$
can be obtained by integrating expression (\ref{d-rho-evap}) over redshift with an account of the drop-off
of the graviton energy density by $(1+z)^{-4}$ and the redshift of the emitted frequency
so that at $t_*=\tau_{BH}$: $\omega= \omega_*  (1+z)$. Note that in the instant decay approximation
the BH temperature remains constant. One has also to take into account that the number density of PBH
behaves as $n_{BH}(t) = n_{p}(t_p)(1+z)^3$, so if we normalize our result to $n_{BH} (\tau_{BH})$, the integrand
should be multiplied by $(1+z)^3$. Finally we obtain:
\be
\frac{\mathrm d\rho_{GW}(\omega_*, \tau_{BH})} {\mathrm d\ln\omega_*} = \frac{0.03 N_{eff}M \omega^4_*}
{ \pi m_{Pl}^4}\, 
(3\tau_{BH})\,{\rho_{BH}(\tau_{BH})}\,I\left(\frac{\omega_*}{T_{BH}}\right)\,,
\label{rho-evap-tau}
\ee
where 
\begin{equation}
I\left(\frac{\omega_*}{T_{BH}}\right)\equiv\varint_{0}^{z_{max}} \frac{ \mathrm dz \left(1+z \right)^{1/2}}
{\exp\left[(z+1) \omega_*/T_{BH} \right] -1}\,,
\end{equation}
and
\begin{equation}
1+z_{max} = \left(\frac{\tau_{BH}}{t_{eq}}\right)^{2/3}\left(\frac{t_{eq}}{t_p}\right)^{1/2}=\left(\frac{32170}{N_{eff}}\right)^{2/3}\left(\frac{M}{m_{Pl}}\right)^{4/3}\Omega_p^{1/3}\,,
\end{equation}
where the effective time of integration is equal to $3\tau_{BH}$ because of the instant decay approximation.
One can check that in this case the total evaporated energy would be equal to the PBH mass.

The spectral density parameter of GWs at $t=\tau_{BH}$ is equal to:
\be
\Omega_{GW}(\omega_*; \tau_{BH})\approx
\frac{ 2.9\cdot 10^3 M^4 \omega_*^4}{\pi\, m_{Pl}^8}\,I\left(\frac{\omega_*}{T_{BH}}\right)\,.
\label{evaporation-BH}
\ee
The spectrum is not a thermal one, though rather similar to it. It has more power at small frequencies due to 
redshift of higher frequencies into lower band and less power at high $\omega_*$. 
The spectral density parameter reaches maximum at $\omega_*^{peak}/T_{BH} = 2.8$. Accordingly the maximum value of the spectral density parameter
when PBHs completely evaporated is equal to:
\begin{equation}
\Omega_{GW}^{peak}(\omega_*^{peak}; \tau_{BH}) \approx 3.8\cdot 10^{-3}\,.
\label{evaporation-BH1}
\end{equation}

\begin{figure}[!htb]
\centering
\includegraphics[scale=.8]{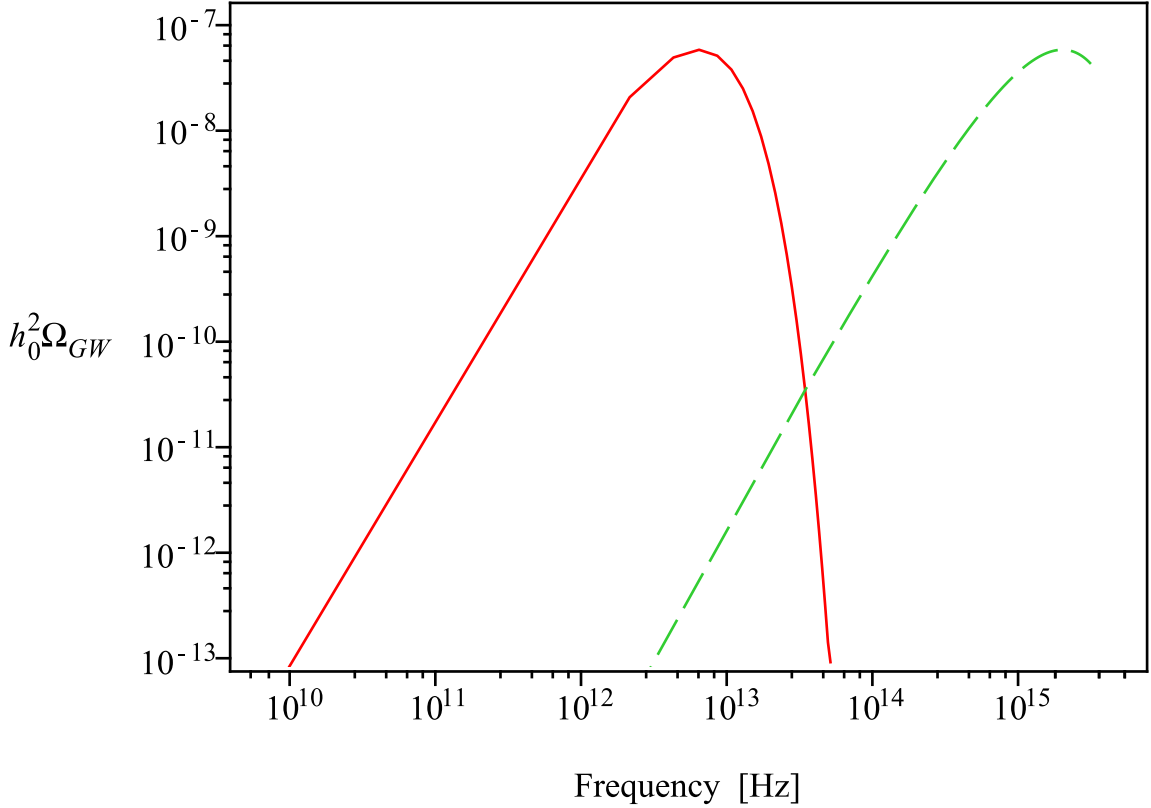}
\caption{Log-log plot of the density parameter per logarithmic frequency, $h_0^2\Omega_{GW}(f; t_0)$, as a function of frequency today, 
$f$, for the case $g_S\sim 100$, $N_{eff}\sim 100$, black hole mass $M=1$ g (solid line) and black hole mass $M=10^{5}$ g (dashed line). 
We can see that the spectrum has a maximum which is sharp and of order $h_0^2\Omega_{GW}(f_{peak})\sim 10^{-7}$.}
\label{fig:evaporation}
\end{figure}

Integrating equation (\ref{evaporation-BH}) first over $\omega_*$ and then over redshift, we find that the total fraction of 
energy of GWs is $0.006$ which is reasonably (in view of the used approximations) close to the expected $0.01$.
At BBN the energy fraction of such GWs would be about 0.005. So the total number of 
additional effective neutrino species would be close to 0.045,
where 0.03 comes from neutrino heating by $e^+e^-$ annihilation and 0.01 comes from the 
plasma corrections (see e.g. review~\cite{nu-rev}). Of course the GWs produced by the considered 
mechanism are safely below the BBN bound~\cite{Abbott:2009ws}.
Using equation \eqref{generalexpression} and taking into account the redshift 
from $t=\tau_{BH}$ to the present time, we find that the total density parameter of GWs today due to PBH evaporation would be about $10^{-7}$.

The total energy density of GWs from the PBH evaporation is quite large but it is concentrated at high frequencies. According to eq. (\ref{omega-0}) the redshifted peak frequency emitted at time $t_*=\tau_{BH}$ becomes today:
\be
f^{(peak)} =  2\cdot 10^{15}\,\textrm{Hz}\,\left(\frac{g_S(T(\tau_{BH}))}{100}\right)^{1/12}\,
\left(\frac{100}{N_{eff}}\right)^{1/2}\left(\frac{M}{10^5\,\textrm{g}}\right)^{1/2}\,.
\label{T_0}
\ee
The energy density of GWs at small $f$ drops down in accordance with equation (\ref{evaporation-BH1}). The spectral density today 
can be calculated from equation (\ref{evaporation-BH}) with an account of the redshift to the present day:
\begin{equation}
h_0^2\Omega_{GW}(f; t_0)=1.36\cdot 10^{-27}\left(\frac{N_{eff}}{100}\right)^2\left(\frac{10^5\,\textrm{g}}{M}\right)^2\left(\frac{f}{10^{10}\,\textrm{Hz}}\right)^4\cdot I\left(\frac{2\pi\cdot f}{T_0}\right)\,,
\end{equation}
where we used $\omega=2\pi f$ and $T_0$ is the BH temperature redshifted to the present time:
\begin{equation}
T_0=\left[\frac{a(\tau_{BH})}{a(t_0)}\right]\,T_{BH}=
4.53\cdot 10^{15}\,\textrm{Hz}\left(\frac{100}{g_S(T(\tau_{BH}))}\right)^{1/12}
\left(\frac{100}{N_{eff}}\right)^{1/2}\left(\frac{M}{10^5\,\textrm{g}}\right)^{1/2}\,.
\end{equation}

\section{ Stochastic background of gravitational waves. An overview  \label{s-GW-background}}

Stochastic background of relic gravitational waves can be produced by several mechanisms. 
The theoretical predictions are model depended due to the uncertainties in the cosmological framework and on 
the values of the redshift from the production epoch. Below we briefly describe some of the production scenarios. For a more detailed review on stochastic background of GWs production mechanisms and their spectra the reader can consult more specific ref. \cite{Allen:1996vm, Maggiore:1999vm, Grishchuk:2007uz}.  

\emph{Inflationary models.} It was established long ago that gravitational waves could be produced in cosmology
due to an amplification of vacuum fluctuations by external gravitational field (quantum particle production).
It was first studied by Grishchuck~\cite{Grishchuk:1974ny} and first applied to an inflationary model by 
Starobinsky\cite{Starobinsky:1979ty}. The gravitational waves could be quite efficiently produced at inflation. Their
spectrum  at large wavelengths is independent on the details of inflationary models.
The frequency band of these gravitons today is quite wide and the associated density parameter is very low. 
The predicted density parameter of gravitational waves in the frequency range from $3\times 10^{-18}$ Hz$<f<$ $10^{-16}$ Hz is 
\begin{equation}
h_0^2\Omega_{GW}(f)\simeq 6.71\cdot  10^{-10}\left(\frac{10^{-18}\textrm{Hz}}{f}\right)^2\left(\frac{H}{10^{15}\textrm{Gev}}\right)^2\,,
\label{inflation1}
\end{equation}
while in the frequency range  $2\times 10^{-15}$ Hz$<f<f_{max}\simeq $ $10^{9}$ Hz the spectrum is flat and the density parameter is 
\begin{equation}
h_0^2\Omega_{GW}(f)\simeq 6.71\cdot 10^{-14}\left(\frac{H}{10^{15}\textrm{Gev}}\right)^2\,,
\label{inflation2}
\end{equation}
where $H$ is the Hubble parameter at inflation.

A near scale-invariant spectrum over a wide range of frequencies is a key prediction of the standard inflationary model \cite{Fabbri:1983us, Abbott:1984fp}. The relative amplitude of GWs spectrum to density perturbations spectrum is usually expressed in terms of the ratio, $r$, of tensor to scalar perturbations. From observations of WMAP, the current limit on B-mode of the CMB polarization demands $r\lesssim 0.22$ which rule out some models of inflation \cite{Komatsu:2008hk, Peiris:2006sj}. The spectrum of GWs can be expressed in terms of the tensorial spectral index, $n_t$, and is almost flat
in the frequency range $2\times 10^{-15}$ Hz $<f<f_{max}\simeq$ $10^{10}$ Hz. The density parameter is proportional to a power
of the frequency:
\begin{equation}
h_0^2\Omega_{GW}(f)\propto f^{n_t}.
\end{equation}
Since the tensorial spectral index is negative, $n_t<0$, the spectrum is decreasing rather than flat. Depending on inflationary model the value of the tensorial spectral index changes and there are some models which predict $r\sim 10^{-3}$.

\emph{Pre-heating phase} at the end of inflation. At this stage the energy of scalar field $\phi$ is spent to 
generate new particles and heat the Universe. The first estimate of the density parameter 
of GWs during the
pre-heating phase was done by Klebnihkov and Tkachev \cite{Khlebnikov:1997di} who found the density 
parameter of the order $h_0^2\Omega_{GW}\sim 10^{-11}$ for the gravitational waves with the present day
frequency $f\sim 10^{6}$ Hz, in the models with quartic potential, $\lambda\phi^4$. Later, this mechanism was 
reconsidered by Easther and Lim \cite{Easther:2006gt, Easther:2007vj} who studied the models with the potentials 
of the form $\lambda\phi^4$ and $m^2\phi^2$. The authors have found numerically 
that   $h_0^2\Omega_{GW}\sim 10^{-10}$ in the frequency range $f\sim 10^8-10^9$ Hz.

\emph{First order phase transitions}. At the end of inflation, first-order phase transitions could have generated a large amount of gravitational waves. 
At such transitions the bubble nucleation of true vacuum states and percolation can occur accompanied by the bubble collisions. 
In a series of papers~ \cite{Turner:1990rc,Kosowsky:1991ua, Kosowsky:1992vn, Kamionkowski:1993fg} 
the energy of gravitational waves generated from bubble collisions at strongly first-order phase transitions 
was estimated and the results were later extended
to the electroweak first-order phase transitions. The amount of GWs from strongly first-order phase transition 
at its end is of the order $1.3\cdot 10^{-3}(\tau/H)$, where $\tau$ is the duration of the phase transition, $H$ is the Hubble 
constant, and the peak frequency is $\omega_*^{peak}=3.8/\tau$. 
The present day  density parameter of GWs produced at the electroweak first-order phase transition 
was found to be of the order $\Omega_{GW}\sim 10^{-22}$ with characteristic 
frequency $f\sim 4\cdot 10^{-3}$. Since later it has been found out, that there is no first order electroweak phase transition 
in the standard model \cite{Kajantie:1995kf}, the mechanism was reconsidered 
by Grojean and Servant \cite{Grojean:2006bp}. The authors estimated the GW production in the temperature 
range 100 GeV-$10^7$ GeV.
The spectrum of the GWs today in this temperature range extends from  $10^{-3}$ Hz to $10^{2}$ Hz. The associated density parameter was found to be quite large, $h_0^2\Omega_{GW}(f_{peak})\sim 10^{-9}$ depending on the parameters of the model.

\emph{Topological defects and cosmic strings.} Practically  in all inflationary models the gravitational wave spectrum 
is almost flat in the frequency range from $10^{-15}$ Hz$<f<f_{max}\simeq $ $10^{10}$ Hz with some variations 
coming from pre-heating and reheating phases for which the frequency is peaked near GHz region. There are other mechanisms 
of GWs production e.g. by cosmic strings which predict almost flat spectrum in a wide range of frequencies. Many of the 
proposed observational tests for the existence of cosmic strings are based on their gravitational 
interactions \cite{Vilenkin:1984ib, Vilenkin:1986hg}. Particularly interesting are GWs produced by closed string 
loops which oscillate in relativistic regime. 
The spectrum of the gravitational waves produced by such relativistic oscillations 
is almost flat in the region $10^{-8}$ Hz$<f<f_{max}\simeq $ $10^{10}$ Hz with a peak at low frequency near 
$f\sim 10^{-12}$ Hz. The density parameter in the frequency range $f\gg 10^{-4}$ Hz, according to ref. \cite{Hogan:2006we}, 
is equal to:
\begin{equation}
h_0^2\Omega_{GW}(f)\simeq 10^{-8}\left(\frac{G\mu}{10^{-8}}\right)^{1/2}\left(\frac{\gamma}{50}\right)^{1/2}\left(\frac{\alpha}{0.1}\right)^{1/2},
\end{equation}
where $G\mu$, $\alpha$ and $\gamma$ are respectively the string tension, the initial loop size as a fraction of the
Hubble radius and the radiation efficiency. From the pulsar timing data the authors of ref. \cite{DePies:2007bm} constrained 
the density parameter of GWs from the cosmic strings in the frequency range $f\gg 10^{-6}$ Hz and put the limit
\begin{equation}
h_0^2\Omega_{GW}(f)\lesssim 10^{-8}. 
\end{equation}
It is generally assumed that at the end of inflation the inflaton oscillates and eventually decays. If non-topological 
solitons, the so called Q-balls, are produced at the inflaton decay, such Q-balls could be a source of GWs. According 
to the calculations of ref. \cite{Mazumdar:2010pn} the density parameter of such GWs would be of the order of 
$h_0^2\Omega_{GW}\sim 10^{-9}$ with a peak frequency $f\sim 10^{10}$ Hz.

\section{Gravitational waves detectors. Present status\label{GW-detectors}}

For most of the models mentioned above, the stochastic background of GWs is beyond the sensitivity of the current and planned interferometers. We have seen that inflationary models predict almost flat spectrum of GWs in a wide range of frequencies. There is a narrow band of frequencies of this background that falls into the range of the present detectors such as LIGO and VIRGO. Unfortunately, the density parameter predicted by inflationary models is too low to be detected by the present detectors. Almost all the models mentioned above predict the density parameter of the order $h_0^2\Omega_{GW}\lesssim 10^{-5}$ and actual LIGO and VIRGO are not able to detect such a quantity because of the
frequency dependence of the density parameter. This can be seen from the relation between the expected 
amplitude of stochastic gravitational waves $h_c(f)$ with the density parameter as 
presented in the ref. \cite{Maggiore:1999vm}
\begin{equation}
h_c(f)=1.3\times 10^{-18}\sqrt{h_0^2\Omega_{GW}(f)}\left(\frac{1 \textrm{Hz}}{f}\right).
\end{equation}
Present detectors such as LIGO and VIRGO with enhanced technologies 
operate in the frequency range 1 Hz - $10^4$ Hz and can reach respectively the 
strain sensitivity $h_{rms}\sim 10^{-23}$ Hz$^{-1/2}$ and  $h_{rms}\sim 10^{-22}$ Hz$^{-1/2}$ in the frequency band $f\sim 10^2 - 10^3$.
\begin{figure}[!htb]
\centering
\includegraphics[scale=.75] {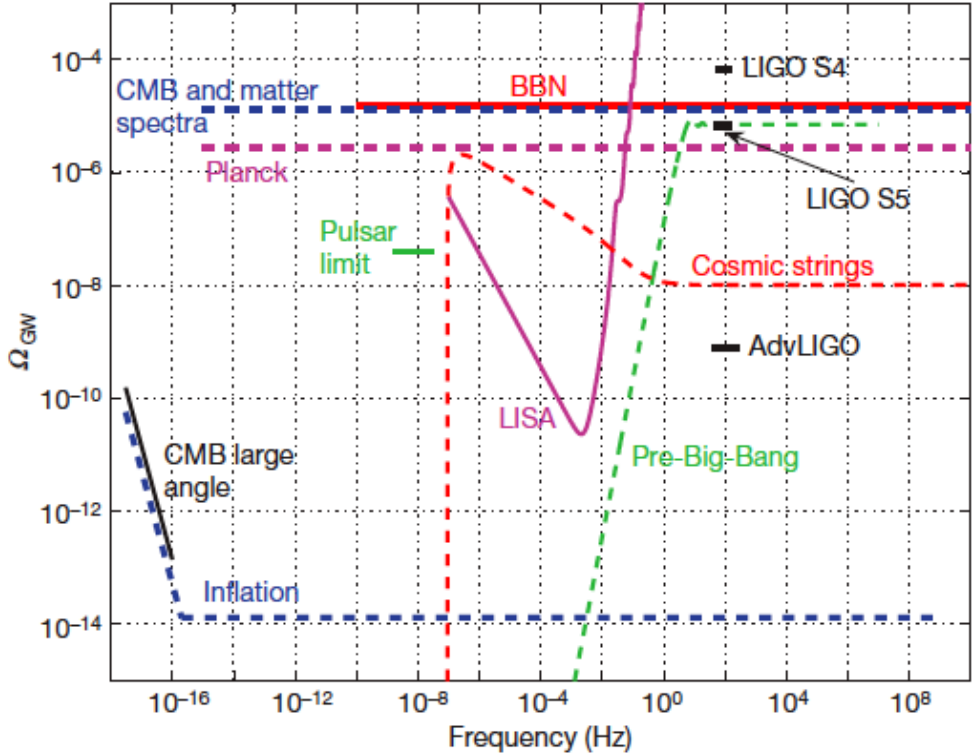}
\caption{$\log[h_0^2\Omega_{GW}$(f)] vs. $\log$(f [Hz]) for different models of production of stochastic background of GWs as given in ref. \cite{Abbott:2009ws}.}
\label{fig:models}
\end{figure}
The planned detectors such as \href{http://www.advancedligo.mit.edu/summary.html}{Advanced LIGO}, \href{https://wwwcascina.virgo.infn.it/advirgo/}{Advanced VIRGO} and \href{http://lisa.nasa.gov/}{LISA} have better chances to detect this stochastic background. In fact, LISA can reach the density parameter of the order $h_0^2\Omega_{GW}\lesssim 10^{-11}$ at frequency $f\sim 10^{-3}$ Hz and Advanced LIGO can reach a  $h_0^2\Omega_{GW}\lesssim 10^{-9}$ at frequency $f\sim 10^{2}$ Hz. These planned detectors can register the stochastic background of GWs coming from cosmic strings and the pre-bing bang stage. The gap between LISA and the ground based detectors will be covered by DECIGO/BBO detectors which will operate in the frequency range from 0.1 Hz to 10 Hz and have $10^3$ better sensitivity than LISA 
from  0.1 Hz to 1 Hz \cite{Takahashi:2003wm}. DECIGO will be able to observe the stochastic background of GWs produced at inflation and can reach $h_0^2\Omega_{GW}\sim 10^{-20}$ at $f\sim 1$ Hz after 3 years of observation \cite{Seto:2001qf, Kawamura:2008zz}. All the above mentioned GWs detectors cover a frequency range $10^{-7}$ Hz - $10^{3}$ Hz and the high-frequency range will hopefully explored by future high-frequency GWs detectors. The principle of a high-frequency detector is based on the electromagnetic-gravitational resonance first proposed by Braginsky and Mensky \cite{Braginsky:1971, Braginsky:1972, Braginsky:1972bz}. Actually there is a  renewed interest on these new detectors which a prototype has been constructed at Birmingham University \cite{Cruise:2000za, Cruise:2005uq, Cruise:2006zt} and which reaches a strain sensitivity of the order $h_{rms}\sim 10^{-14}$ Hz$^{-1/2}$ at $f\sim 10^8$ Hz. The main goal of this detector is detection of high-frequency stochastic background of GWs from the early Universe and black hole interactions in higher dimensional gravitational theories.

\section{Summary and results\label{conclusion}}

We have analyzed the formation and evolution of light primordial black holes in the early Universe which created a transient matter domination regime in contrast to the present standard cosmology, where the early Universe after inflation was normally radiation dominated. PBHs with masses less than $M\sim 10^8$ g evaporated before primordial nucleosynthesis leaving no trace. Thus the fraction of the energy density of such PBHs, $\Omega_p$, in this case is a free parameter of the model, not constrained by any existing observations. 

At MD stage the PBHs could form high density clusters which 
would be efficient sources of the primordial GWs. PBHs could have dominated the Universe for a short time of the order of their lifetime, 
$\tau_{BH}$, generating relic gravitational waves by various mechanisms of their mutual interactions 
as well as due to their evaporation. In the former case we have shown that production of GWs is most efficient after BH density started to dominate over radiation. After that moment, high density clusters of PBHs could have been formed, leading to an efficient production of GWs.
To survive till cluster formation the PBH mass at production must be bounded from below by  $M\sim 4\cdot 10^{-5}$ g $\Omega_p^{-1}\Delta_{in}^{-3/4} N_{eff}^{1/2}$ which leads to a lower bound $\Omega_p>10^{-14}\Delta_{in}^{-3/4} N_{eff}^{1/2}$. According to the standard cosmology the amplitude of primordial density perturbations is of order of $\sim 10^{-4}$, which in our case leads to a lower bound on the density parameter of PBHs, $\Omega_p\gtrsim 10^{-11}$.

In this context we have calculated the density parameter of GWs today from scattering of PBHs in both classical and quantum regime, GWs emission 
from binaries, and from black hole evaporation. We have shown that a substantial amount of gravitational waves has been emitted by all mechanism considered here. In the case of scattering of PBHs we considered \emph{only} scattering between them neglecting the possibility of PBHs mergers, which results in an underestimate on $h_0^2\Omega_{GW}$. Even in this case the density parameter is substantial at high frequencies reaching values of the order of $h_0^2\Omega_{GW}\sim 10^{-9}$ at $f\sim$ GHz for classical scattering and 
the total density parameter $h_0^2\Omega_{GW}\sim 10^{-10}$ for very light primordial black holes. In the low frequency limit the density parameter in the classical case is of the order of $h_0^2\Omega_{GW}\sim 10^{-17}-10^{-20}$ in the frequency range $f\sim 10^{-1}-10^{2}$ Hz which falls into the detection band of DECIGO/BBO. 

The number of PBHs that form binaries after cluster formation is subjected to uncertainties and in this paper we parametrized it through factor $\epsilon$. The exact value of this parameter could be calculated elsewhere by numerical calculations. Since the density in such clusters is very high we expect that $\epsilon$ is not very small in comparison with unity. In fig.\ref{fig:binaries} the expected value of the density parameter today is presented. We can see that a large amount of gravitational waves has been emitted in the high frequency regime with $h_0^2\Omega_{GW}\sim 10^{-14}- 10^{-12}$ at frequency $f\sim 10^{10}$ Hz depending on the BH initial mass. In the low frequency part of the spectrum the spectral density parameter is utterly negligible making it impossible to detect GWs produced by this mechanism at present and probably in the near future. In our derivation we have considered both stationary and inspiral phases of binaries leading to a wide range of the frequencies emitted. We have considered only binaries in circular orbits and the problem with elliptical orbits will be treated later. 
If elliptical orbits were frequent, 
the amount of GWs will be presumably higher over a wide range of frequencies. We assumed that all binaries are formed with initial radius less than the average distance between PBHs and greater than the gravitational radius $r_g$. In this case the frequency spectrum has a cutoff in both low and high frequency bands of the spectrum.

Another mechanism of graviton productions considered here is the PBHs evaporation. This mechanism is independent on the structure 
formation during the PBH domination. In fig.\ref{fig:evaporation} we show the density parameter as a function of frequency for BH masses $1$ g and $10^5$ g. Having a near blackbody spectrum, the frequency of the emitted gravitons 
can have any value, but unfortunately the GWs spectrum has a peak in the high frequency region which today make a substantial contribution into the cosmological energy density of the order of $h_0^2\Omega_{GW}(f_{peak})\sim 10^{-7}$.

The mechanisms considered in this paper could create a rather high cosmological fraction of the energy density of the relic gravitational waves at very high frequencies and gives an opportunity on investigating the high GWs spectrum by present and future detectors. Unfortunately at the lower part of the spectrum $\Omega_{GW}$ significantly drops down. Still the planned interferometers DECIGO/BBO could be sensitive to the predicted GWs. It is noteworthy that the mechanism of GWs generation suggested here kills or noticeably diminishes GWs from inflation by the redshift of the earlier generated GWs at the PBH (MD) stage.

\end{document}